\documentclass[]{aa}
\usepackage{natbib}
\usepackage{graphicx}
\usepackage{subfigure}
\usepackage{captcont}

\begin{document}

\title{Pre-main sequence stars in the stellar association N11 in the Large Magellanic Cloud}
\subtitle{}

\author{Antonella Vallenari \inst{1}, Emanuela Chiosi\inst{2}, Rosanna Sordo \inst{1} }

   \institute{INAF, Padova Observatory,
Vicolo dell'Osservatorio 5, 35122 Padova, Italy     \\
\and Astronomy Department, Padova University, Vicolo
          dell'Osservatorio 2, 35122 Padova, Italy \\
   \email{antonella.vallenari@oapd.inaf.it; emanuela.chiosi@oapd.inaf.it; rosanna.sordo@oapd.inaf.it }
             }

\offprints{A. Vallenari}

\date{Received: August 2009 / Accepted:  }

\abstract{Magellanic Clouds are of extreme importance to the study of the star formation process in low metallicity environments. }{In this paper we report on the discovery of  pre-main sequence candidates and young embedded stellar objects in N11 located in the Large Magellanic Cloud to cast light on the star formation scenario. We would like to remind that this comparison is complicated by the presence of  a large age dispersion detected in the fields.}{Deep archive HST/ACS photometry is used to derive color-magnitude diagrams of the associations in N~11 and of the  foreground field population. These data are complemented by archive  IR Spitzer data which allow the detection  of young embedded stellar objects. The spatial distribution of  the pre-main sequence candidates and young embedded stellar objects is compared with literature data observed at different wavelengths, such as  H$_{\alpha}$ and CO maps, and with the distribution of OB and Herbig Ae/Be stars. The degree of clustering is derived using the Minimal Spanning Tree method and the two point correlation function to get insights about the formation process.} {A large population of pre-main sequence candidates is found in N~11. Their masses are in the  range of 1.3$-$2 M$_\odot$ for ages from 2 to 10 Myr.  Young embedded stellar objects having ages of 0.1$-$1 Myr are found to be intermixed with the candidate pre-main sequence stars. The spatial distribution of the stars shows that this region is the product of clustered star formation. No significant difference is found in the clustering degree of young blue main sequence stars  and faint pre-main sequence candidates, suggesting that they might be part of the same formation process.}
 {The data  suggest that the star formation in the region is a long-lasting process where stars from 0.1 to 10 Myr are widely distributed.}

\titlerunning{Pre-Main Sequence stars in N~11 in the LMC}
\authorrunning{A. Vallenari et al.}

\maketitle

\section{Introduction}\label{sec_intro}

Several plausible scenarios are presented in the literature to explain how isolated stars form  \citep[e.g][]{2004MNRAS.349..735B, 2003ApJ...585..850M}. A detailed understanding of how a molecular cloud is converted into stars is still missing  \citep[see among others] []{2000ApJ...530..277E, 2007ApJ...661..262D, 2007ApJ...665..478K}.
Interstellar turbulence is found to be  very efficient in sweeping up molecular gas-forming massive structures which in turn can undergo a large-scale gravitational collapse. Supersonic turbulence can counterbalance gravity on a global scale, but provokes collapse locally \citep{1973PASJ...25....1S, 2003IAUS..208...61K, 2003ASPC..287...65L}.
As a consequence, clusters are  formed in a hierarchical fashion with subclusters which eventually merge to build up the final condensed object \citep{2009ApJ...694..367S}. Stellar feedback is expected to strongly influence the process.
Young massive stars inject energy into the nearby interstellar medium, heating and compressing the surrounding gas.  This  process can have destructive or constructive effects, depending on the balance between heating and gravity, but it is still not clear  what regulates it.
 The presence or absence of turbulent feedback directly relates to the physical mechanism of star formation and determines whether stars are generated by the formation and collapse of discrete protostellar cores \citep{2005Natur.438..332K, 2002ApJ...576..870P} or by competitive accretion \citep{2004MNRAS.349..735B}.

To analyze entire star-forming regions including  different components (molecular clouds, ionized gas, stars.. ) can cast light on the formation scenario. Clustering properties of the young population can give information about the structure of the interstellar medium from which it formed.
The Magellanic Clouds are ideal laboratories to study the process of star formation in detail, due to their proximity. In the Large Magellanic Cloud (LMC) several regions of active star
formation are found.
N11 is the second largest nebula of the LMC after the 30 Doradus
Nebula. It is located at the  North-Western corner of the LMC.

 This region  is often presented as one of the best examples of triggered star formation \citep{1992ApJ...399L..87W, 1996A&A...308..588R, 2006AJ....132.2653H, 2007A&A...465.1003M}. The basic idea is that the association LH~9 has triggered star formation in LH~10 and LH~13 (see discussion in the following).

Here we discuss the stellar content in the region of N~11 using  HST ACS/WFC archive data.
These observations are complemented by Spitzer archive data: while HST observations can give information about faint, exposed pre-main sequence candidates (PMSs), IR data allow to detect embedded young stellar objects(YSOs).
Regions of  active star formation can be detected, pointing out the  differences between the areas where the nebulae are located and the surrounding fields.\\
This paper is part of a project aimed to cast light on the process of field and cluster star formation in the Magellanic Clouds. We quote among others \citet{2006A&A...452..179C} where the  cluster and field star formation in the central part of the Small Magellanic Cloud (SMC)  is  studied and  the correlation between young objects and their interstellar environment is discussed. In \citet{2007A&A...466..165C} the star formation rate in the SMC is found to be  quiescent at ages older than 6 Gyr.

This paper is organized as follows: in Sect.~\ref{clu} the cluster and association content of the area is presented; in Sect.~\ref{acs_data} the  HST ACS/WFC and Spitzer archive data are presented;
in Sect~\ref{method} the methods used to analyze the region are discussed; in Sect~\ref{field_pop} the field population is studied and the star formation history is derived. In Sect~\ref{interstellar} an estimate of the interstellar extinction is given;
in Sect~\ref{ind_cmd} the CMDs of the most relevant associations and clusters are presented and their stellar contents and ages are discussed;
in Sect~\ref{clustering} the degree of clustering of the stellar populations at different ages is derived. Concluding remarks are drawn in Sect~\ref{conclu}.

\section{Presenting the region: clusters and associations in N11}
\label{clu}

This section summarizes literature information about the cluster and associations in N11.
The  complex  has a diameter of 45$^\prime$, corresponding to a linear
extent of 655 pc, if  the distance modulus $(M-m)_o=18.5$ \citep{2006ApJ...642..834K}.
This region has a peculiar morphology: a central hole with no emission is
surrounded by several bright nebulae and filaments. 
These structures are observed at the H$\alpha$ and [OIII] 5007 \AA\ wavelengths and in the radio  \citep{1996A&A...308..588R, 2003A&A...401...99I}.\\
The different emission areas are classified according to the emission brightness starting from N11~A
to B,C,D,E,F and I.  The central cavity has been evacuated by the OB stars in the association LH~9, located near the center of N11. The  OB association LH~10 in N11B is to the north of LH~9, while LH~13 in N11C is to the east of LH~9. LH~14 in N11E is situated to the northeast of LH~10 and is outside  the studied region.

\begin{figure}
  \centering
  \resizebox{\hsize}{!}{\includegraphics{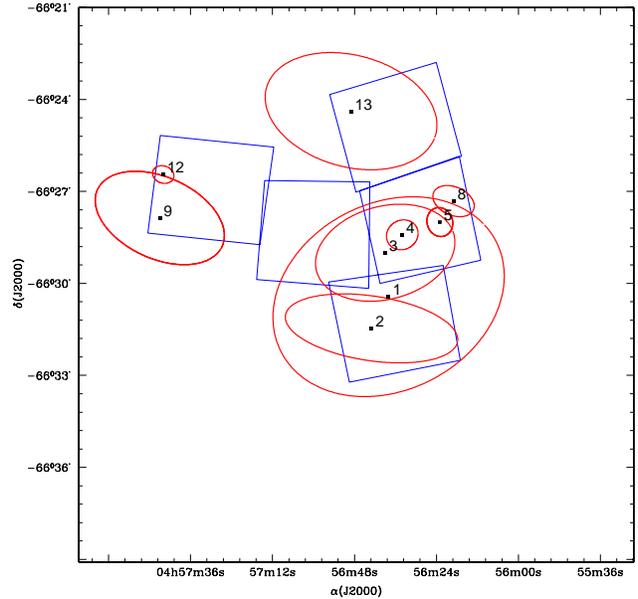}}
 \caption{The region containing the studied HST/ACS fields (boxes) is schematically
 represented.  The ellipses indicate objects of particular interest, as
 clusters and associations.  The numbers refer to the object ID listed in Table \ref{tab1}.  }
   \label{n11clu}
    \end{figure}

The \citet{1999AJ....117..238B} Catalog lists 49 clusters and associations in the whole region of N11. This catalog collects and cross-identifies stellar and non-stellar objects (nebulae, supernova remnants) out of more than 30 catalogs. In our observed region, 13 objects are detected.
Table \ref{tab1} lists the associations and clusters in the
observed region from the catalog of \citet{1999AJ....117..238B} together with their minimum and maximum diameter, and the \citet{1999AJ....117..238B} classification. For clarity, we recall it. The sequence C~(Clusters) to A~(Associations) refers to the density, where high density objects  are classified as C, and low density objects as A. Objects NA and NC are stellar systems clearly related to emission. Figure~\ref{n11clu} presents a map with the location of clusters and associations of interest.
\cite{2006AJ....132.2653H} find 127 Herbig Ae/Be star candidates  from near-infrared photometry, mainly in the periphery of LH~9. Herbig Ae/Be star are intermediate mass pre-main sequence stars (3$-$7 M$_\odot$) having an age range of 1$-$3 Myr.

\begin{table*}
\caption{Clusters and associations of particular interest in the
region under investigation.
Column 1 gives the identifier(ID), Cols.~2 and 3 the coordinates (J2000), Col.~4 gives the classification by Bica et al, Cols.~5 and 6 give the minimum and the maximum values of the diameter, Cols.~7 and 8 list the name and the location.   } \tiny
\begin{center}
\begin{tabular}{|l | l | l | l| r | r | l| l | l }
\hline
ID  &  Ra & Deg & Class & D& D      & Name & Location  \\
&&&&(arcmin)&(arcmin)&&\\
\hline
    &       &     &       &  Max   & Min    &      &      \\
 \hline
1  &   04 56 38 & -66 30 26& NA  &     7.30 &  5.90  &  LH~9                  &   LMC-N11 \\
2  &   04 56 43 & -66 31 28& NA  &     5.10 &  2.10  &  NGC~1760,             &   LH~9     \\
   &            &          &     &          &        &  LMC-N11F,            &              \\
   &            &          &     &          &        &  ESO85EN19            &              \\
3  &   04 56 39 & -66 29 00&  A  &     4.20 &  3.00  &  NGC~1761,             &   LH~9     \\
   &            &          &     &          &        &  SL~122,               &              \\
   &            &          &     &          &        &  ESO85SC18            &              \\
4  &   04 56 34 & -66 28 25&  C  &     1.00 &  0.90  &  HD~32228,             &   NGC~1761 \\
   &            &          &     &          &        &  KMHK~307              &              \\
5  &   04 56 23 & -66 28 00& NA  &     0.95 &  0.75  &  BSDL~270              &   NGC~1761 \\
6  &   04 56 30 & -66 26 40& NA  &    26.00 & 23.00  &  LMC-N11,             &   LMC-N10 \\
   &            &          &     &          &        &  LMC-DEM34            &              \\
7  &   04 57 44 & -66 28 30& NC  &     0.30 &  0.25  &  HNT1                 &   NGC~1769 \\
8  &   04 56 19 & -66 27 19&  A  &     1.30 &  0.90  &  BCDSP~1               &   LH~9     \\
9  &   04 57 45 & -66 27 52& NA  &     4.10 &  2.60  &  NGC~1769,LH~13,        &   LMC-N11 \\
   &            &          &     &          &        &  LMC-N11C             &              \\
   &            &          &     &          &        &  ESO85SC23            &              \\
10 &   04 56 09 & -66 31 25& NA  &     1.50 &  0.90  &  BSDL~264              &   LH9     \\
11 &   04 57 17 & -66 24 56& NA  &     0.80 &  0.70  &  HT3                  &   LMC-N11 \\
12 &   04 57 44 & -66 26 27& NA  &     0.65 &  0.55  &  BSDL~324              &   NGC~1769 \\
13 &   04 56 49 & -66 24 23& NA  &     5.20 &  3.60  &  NGC~1763,             &   LMC-N11 \\
   &            &          &     &          &        &  IC~2115,              &              \\
   &            &          &     &          &        &  LMC-N11B,            &              \\
   &            &          &     &          &        &  SL~125, LH~10          &              \\
\hline
\end{tabular}
\end{center}
\label{tab1} \normalsize
\end{table*}

\section {The data}
\label{acs_data}

Here we first present the HST ACS/WFC archive data we use together with the data reduction and calibration. Then  we  discuss the infrared Spitzer archive data.

\subsection{HST ACS/WFC photometry of the stellar population}

Data for six fields in N11 were taken from the ACS/WFI HST archive (PI Ma\'iz-Apell\'aniz, Proposal ID=9419). The studied fields are  partially covering the associations LH~9, LH~10, and LH~13.
In Fig.~\ref{n11clu} we sketch the position of the ACS fields in the N11
complex.
The control field to sample the background and foreground population is located outside the N~11 boundaries and is not plotted.
Observations are made in the filters F435W and F814W with  long exposures of 496 s and  short exposures of 1$-$2 s. We make use of drizzled files which already account for image distortion.  The data are then reduced by means of the packages DAOPHOT/ALLSTAR by \citet{stetson1994}.

The correction for  charge transfer efficiency  is applied following \citet{sirianni2005}.
The instrumental magnitudes are transformed to the ACS-Vega system using the zero points derived by
\citet{bohlin2007}.

\begin{figure}[h]
\parbox{8cm}{
\resizebox{8cm}{!}{\includegraphics{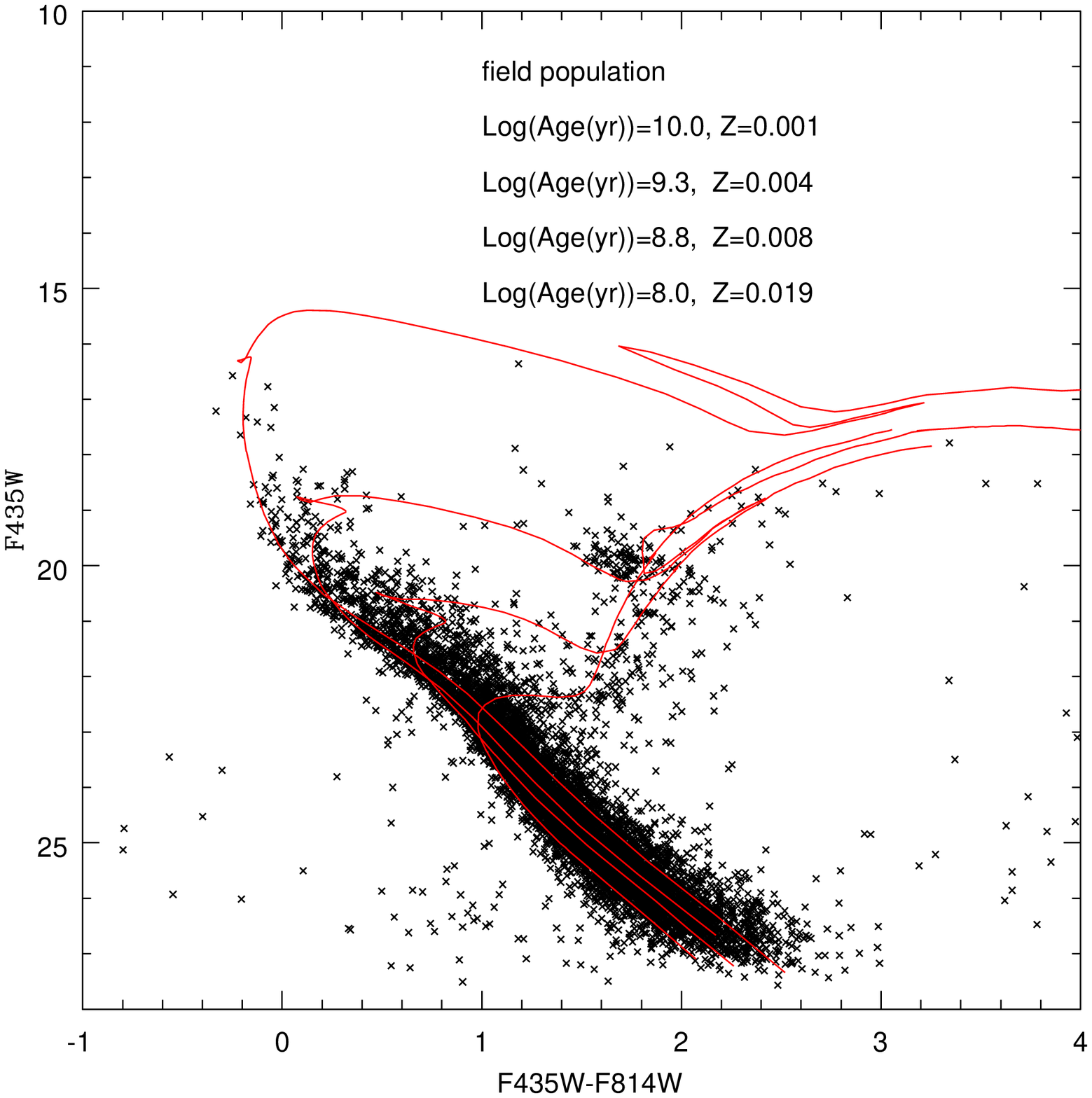}}\\
}
\parbox{8cm}{
\resizebox{8cm}{!}{\includegraphics{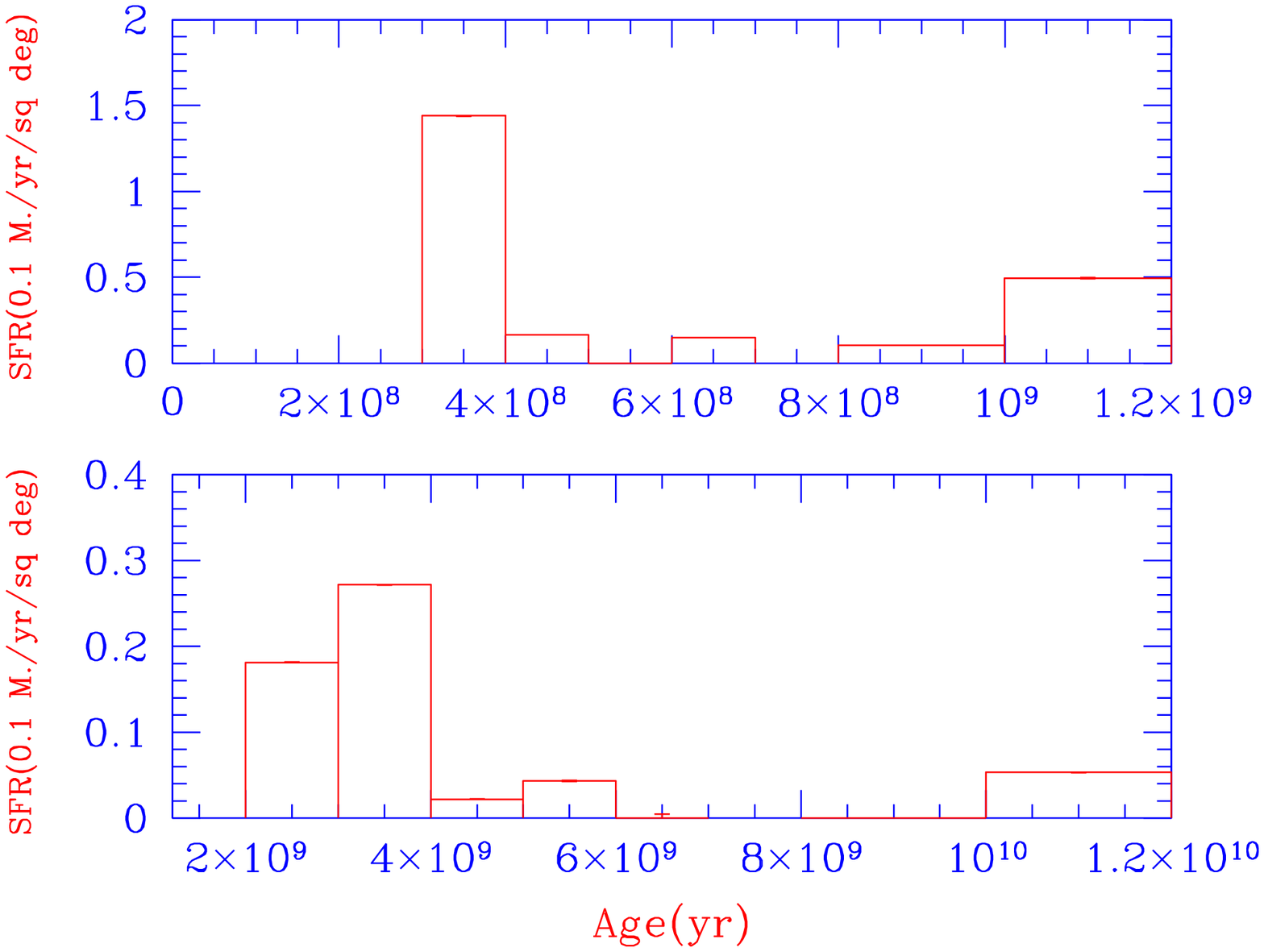}}\\
}
 \caption{CMD of the field population where isochrones of different ages and chemical composition  are superposed, namely Age=0.1 Gyr, Z=0.019; Age=0.6, Gyr Z=0.008; Age=2 Gyr, Z=0.004; Age=10.0 Z=0.001 Gyr(upper panel).
In the lower panel, the SFR in the field population is presented as a function of age (in yr) (see text for details).}
\label{sfr_field}
\end{figure}

Figure~\ref{sfr_field} presents the CMD  of the background population.

\begin{figure}
   \centering
  \resizebox{7cm}{!}{\includegraphics{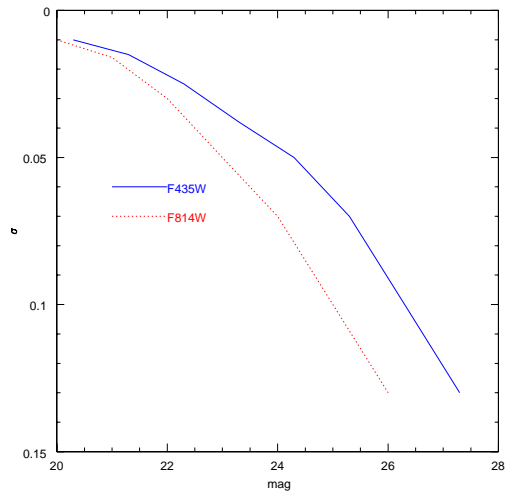}}
 \caption{Photometric errors in the two pass-bands as a function
 of the magnitude, as recovered from artificial star experiments. }
   \label{n11_phot_error}
    \end{figure}

\textbf{Photometric errors}. Photometric errors are estimated in the usual way by means of  artificial star experiments, which consists of the injection  of a large number of artificial stars of known magnitude into the original image: stars are then recovered through the whole reduction pipeline, and recovered magnitudes are compared to the original ones. The mean magnitude  difference between injected and recovered stars, which is taken as an estimate of the mean photometric error, is plotted in Fig.~\ref{n11_phot_error} as a function of the magnitude.\\

\textbf{Completeness factor.} The completeness factors
$\Lambda$ are calculated  for each field from crowding tests as the ratio between the
recovered stars in a given magnitude interval and the number of
original stars in the same interval.

{   In the field population region the completeness is 90\% at F435W$\sim 26.0$ and F814W$\sim 25.0$. 
In the associations, the mean completeness is  better than 90\%  for  F814W$<$ 24.5 and F435W $<25.0$ mag, it becomes 80\% at F814W$\sim$ 25.0 and F435W $\sim 25.5$ mag, and then steeply declines. At a changing field, the completeness  varies by about 7\% around the mean value, being slightly worse for the field including LH~13. In the following, the analysis of the luminosity functions (LFs) is limited to  F435W $< 25.5$ mag (or  F435W$_o < 24.5$ mag), where the completeness correction is quite negligible.}

\subsection{Infrared Spitzer data}

The region was observed by the infrared survey  SAGE   (Surveying the Agents of a Galaxy's Evolution) using the Infrared Array Camera IRAC (3.6, 4.5, 5.8 and 8 $\mu$m) and MIPS (24 $\mu$m) instruments on board of the Spitzer Space Telescope \citep{2006AJ....132.2268M, 2007IAUS..241..363M}.
All relevant information concerning those archive data, data reduction, and the catalog can be found in \cite{2006AJ....132.2268M, 2007IAUS..241..363M}.
Infrared CMDs and color-color plots will be discussed in the following sections.

\section{The methods}
\label{method}
This section discusses the methods used to study the stellar population in N~11.
First, we summarize the Downhill Simplex procedure  used to derive the star formation history (SFR) of the field population. Second, we describe how we subtract the field star contamination from HST CMDs of the associations  and how we derive the extinction. Finally, we discuss how the two-point correlation function is used to derive the degree of clustering of the stellar population.

\subsection{Field population SFR: Downhill Simplex method}
 
A detailed description of the use of the Downhill Simplex  method to derive the SFH of the field population can be found in \cite{chiosi2006}, \cite{2007A&A...466..165C}. Here we recall a few points:

(1) To infer the SFR, theoretical CMDs in different age ranges are simulated. The simulations include the spread which is due to the observational photometric errors, reddening, and the effect of photometric incompleteness. In each age bin,  10000 to 15000 synthetic stars are generated down to the  photometric completeness limit set at the magnitude where the incompleteness factor is 50\%.
The synthetic CMDs stand on the sets of stellar tracks, isochrones and single stellar populations by \cite{girardi2002}. We assume the chemical enrichment history in agreement with \cite{pagel1999}, \cite{2008AJ....135..836C}. Table \ref{pop.tab} lists the ages and metallicities of the synthetic stellar populations used in the simulations.

(2) The extinction along the line of sight of the  field population is estimated by means of the fit of the CMD with a set of synthetic stellar populations.

(3) The relative contribution of different populations to the total CMD and luminosity/color functions to be compared with the observational ones, in other words the SFR, is derived from a minimization algorithm which makes use of the Downhill Simplex method \citep{harris2004}.

(4) To provide the required constraints to the minimization procedure, we split the  observational CMD into a number of suitable magnitude-color bins. Recent work concerning the determination of the SFR from the CMDs has pointed out the importance of using a binning that takes the various stellar evolutionary phases into account, as well as the uncertainties of the stellar models \citep{2002ASPC..274..490R,2006A&A...452..179C}.  To avoid spurious solutions, the fit is mainly based on the main sequence magnitude. To properly sample the age of the population, the magnitude bin  is variable, being 0.5 mag from 16 to 19 mag, and 0.25 mag from 19 mag to 25 mag.

\begin{table}
\caption{Ages and metallicities of the synthetic populations in
use.}
\begin{center}
\begin{tabular}{ c   c }
\hline
\multicolumn{1}{c}{Age (Gyr)}  & Z  \\
\hline
 0.08  $-$  0.12   & 0.006 $-$ 0.010 \\
 0.12  $-$  0.30   & 0.006 $-$ 0.010 \\
 0.30  $-$  0.40   & 0.006 $-$ 0.010 \\
 0.40  $-$  0.50   & 0.006 $-$ 0.010 \\
 0.50  $-$  0.60   & 0.006 $-$ 0.010 \\
 0.60  $-$  0.80   & 0.006 $-$ 0.010 \\
 0.80  $-$  1.00   & 0.006 $-$ 0.010 \\
 1.00  $-$  2.00   & 0.006 $-$ 0.010 \\
 2.00  $-$  3.00   & 0.005 $-$ 0.010 \\
 3.00  $-$  4.00   & 0.003 $-$ 0.005 \\
 4.00  $-$  5.00   & 0.003 $-$ 0.005 \\
 5.00  $-$  6.00   & 0.003 $-$ 0.005 \\
 6.00  $-$  8.00   & 0.0017 $-$ 0.003~ \\
~8.00  $-$  10.00  & 0.0017 $-$ 0.0017 \\
10.00  $-$  12.00  & 0.0017 $-$ 0.0017 \\
\hline
\end{tabular}
\end{center}
\label{pop.tab}
\end{table}

\subsection{Field star subtraction}
\label{field_sub}

Field subtraction is a critical issue to isolate the population of N~11.
Field stars are statistically subtracted from the CMDs of the association. 
First, the CMDs of both association and field are divided in boxes of the size of $\Delta F435W=0.2$ and $\Delta (F435W-F814W)=0.3$. The incompleteness correction is taken into account by dividing the field and association CMDs in magnitude-color bins and then adding on each bin  having N$_{th}$ stars,
$(1-\Lambda)\times N_{th}$ objects, where $\Lambda$ is the smallest of the $F435W$ and
$F814W$ completeness factors. {At each CCD area in both the association and in the field the appropriate completeness correction as calculated from artificial star experiments is applied. Then the closest cluster  star is subtracted in each box of the N11 CMD for every field star.}
This procedure was already used in a number of papers where more detail can be found \citep{2006A&A...452..179C, 2007A&A...466..165C}.

\subsection{Determination of the interstellar extinction in N~11}

In order to study the stellar population in the association, it is necessary  to derive a detailed map of the interstellar extinction which is expected to be highly variable.
First, we select the main sequence stars brighter than
F435W $\sim 18$.  The comparison between the CMDs of the field population and those of N11 shows that these stars have a high probability to belong to the association, since they are brighter than the youngest turnoff of the field population (see Figs. \ref{cmd_1.fig} and \ref{sfr_field}). About 1000 stars are found in this magnitude range. Then, assuming that the spread in color of the stars at a given magnitude is due to interstellar extinction only,  their color $(F435W-F814W)$ is compared with the intrinsic color  expected for main sequence stars having an age of 5 Myr (see discussion in Sect.~6), which allows the determination of the color excess E${(F435W-F814W)}$ for each star.
We assume  the extinction law $A_{F435W}/A_{F814W}$ by  Sirianni et al~(2005). Then the  E${(F435W-F814W)}$ values are interpolated using the package GRIDDATA in IDL as a function of the coordinates.
Finally, the extinction of each star in the field is calculated from the interpolated map.

\subsection{Statistical analysis of the stellar spatial distribution} 

We present the methods used to study the degree of clustering of the stars and derive the spatial scale of the formation. First, we define the two-point correlation function, second we describe  the minimal spanning tree method.

\subsubsection{Two-Point correlation function}

The probability $1+\xi(r)$ of finding a neighbor in a shell element at a distance r from an object of the sample is

\begin{equation}
1+\xi(r)= 1/(Nn) \sum_{i=1}^N n_i(r),
\end{equation}

where $\xi(r)$ is the two-point correlation function,  $n_i(r)$ is the number density of objects found in an annulus centered on the i-th object and having a radius between r and r+dr, N is the total number of objects, and finally, $n$ is the average number \citep{peebles1980}.  A Monte Carlo algorithm is used to derive the area covered by the data when the annulus extends outside the studied region.
Using the above definitions, a random distribution of stars will produce a flat correlation, with $\xi(r) \sim 0$.  A peaked $\xi(r)$ at small radii  indicates a positive correlation, and the full width half maximum of the peak itself represents the spatial scale of the clustering. The distribution  of the nearest neighbor distance might reflect the structure of the interstellar medium from which the stars formed. The absolute value of $\xi(r)$ is a measure of the concentration of the objects at a given distance r  relative to the average distribution \citep{peebles1980}.  

\subsubsection{Minimal Spanning Tree}

The Minimal Spanning Tree (MST) is a powerful method used in literature for a statistical analysis of the spatial distribution of the stars in a cluster 
\citep[see][for a detailed discussion]{2004MNRAS.348..589C, 2006A&A...449..151S}.  The MST is the unique network of straight lines joining a set of points (vertexes), so that it  minimizes the total length of all the lines (called edges) without creating closed loops.
In this paper, we make use of the routine {\it mst} \citep{ACM-TRANS}, based on Prim's algorithm \citep{Prim1957} to construct the MST. Starting from an arbitrarily chosen first point, an edge is created joining that point to its nearest neighbor. The tree is then calculated constructing the shortest link between one of its nodes and the unconnected points, until all the points are connected.
Once a MST is calculated,  its   mean edge length  $\bar{m}$ is directly defined.  The  $\bar{m}$ is dependent on the total number $N$ of stars in the cluster, in the sense that as $N$ increases, shorter edges are created and $\bar{m}$ decreases. For this reason, \citet{2004MNRAS.348..589C} normalize $\bar{m}$ to  $(\textsl{N}A)^{1/2}/(\textsl{N}-1)$  when comparing clusters that have a  different object number $N$ and/or areas $A$. \citet{2004MNRAS.348..589C} and \citet{2006A&A...449..151S} demonstrate that the structure of a cluster can be derived comparing the $\bar{m}$ of the MST describing it, to the normalized correlation length $\bar{s}$  defined as the mean separation between the stars  normalized to the radius of the region.

This defines the parameter $\textsl{Q}$ as 
$\textsl{Q}=\frac{\bar{m}}{\bar{s}}$.

\citet{2004MNRAS.348..589C} show that  $\bar{s}$ decreases more quickly than  $\bar{m}$ as the degree of the central concentration becomes more severe (indicating the presence of a radial gradient), while  $\bar{m}$ decreases more quickly than  $\bar{s}$ as the degree of subclustering is getting higher.
 $\textsl{Q}$ can be used to distinguish between subclustering ( $\textsl{Q}<0.8$) and the presence of  a large scale radial density gradient
 ($\textsl{Q}>0.8$).  It should be remembered that  $\textsl{Q}$ cannot give any information about the fractal nature of the sub-clustering,  i.e. it cannot say whether the subclustering is hierarchically self-similar.

\section{Results: the SFR history of  the field population}

\label{field_pop}

Fig.~\ref{sfr_field} presents the CMD of the field population where isochrones of different ages are superposed.
The youngest field population shows a turnoff at about F435W $\sim$ 17$-$19,
and the stars burning He in the core are located at
F435W $\sim 20$.  The CMD suggests the presence of an intermediate age and young population.

The detailed determination of the star formation rate is presented in Fig.~\ref{sfr_field} (lower panel) and reveals the presence of a small component of a very old population (10$-$12 Gyr), an intermediate age component in the range 2$-$6 Gyr, and several bursts of a younger population, up to 0.4 Gyr. This result agrees with previous determinations by
\cite{dolphin2001, harris2004,
 javiel2005}.

\section{Results: the interstellar extinction}
\label{interstellar}

\begin{figure}[th]
\resizebox{8cm}{!}{\centering {\includegraphics[width=8truecm,height=8truecm]{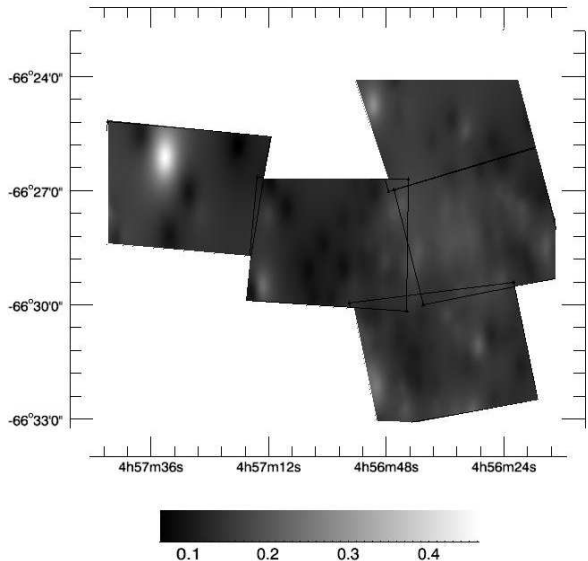}}}

\vspace{1.truecm} 
\caption{Extinction map in E(F435W-F814W).}

\label{imaext}
\end{figure}

The map of the interstellar extinction in E$(F435W -F814W)$ is presented in Fig.~\ref {imaext}.
The interstellar extinction E(F435W-F814W) is found to be in the range of 0.2$-$0.5. This corresponds with A$_{F435W} \sim 0.95$-$0.38$ using the relations by \citet{sirianni2005} and agrees with the values derived by means of the spectroscopy of bright stars by \cite{2007A&A...465.1003M}.
The main source of uncertainty  on our determination of the extinction is due to the presence of an age spread inside the young main sequence stars.
We derive the extinction using main sequence isochrone with an age of 5~Myr, while ages as young as 2~Myr are found for the blue population (see discussion in the following Section). At F435W=12.5, the expected  MS color difference between an age of 2 Myr and 5 Myr is of 0.07 mag, while at F435W=14 is 0.01 mag.  These values can be regarded as an estimate of the uncertainty on E$(F435W -F814W)$.

\begin{figure}[htp]
\parbox{8cm}{
\resizebox{8cm}{!}{\includegraphics{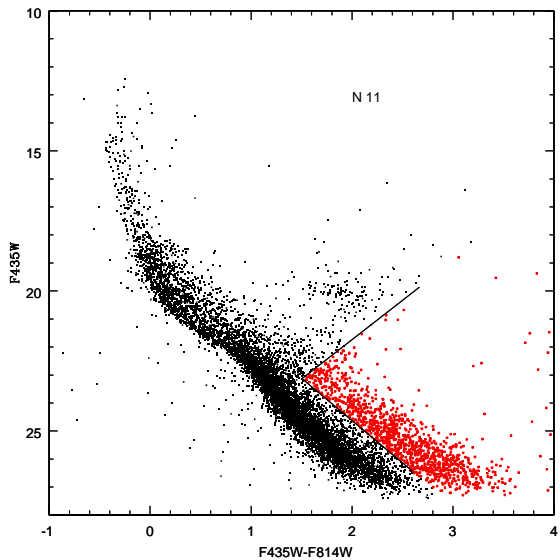}}\\
}

 \caption{{   CMD of the N~11 region covered by the HST/ACS photometry.} The lines drawn in the CMD mark the regions redder than the MS where  PMSs are clearly separated from the field population.  These lines
indicate  only a part of the region in the diagram where PMSs
are located.}
\label{cmd_1.fig}
\end{figure}
 
\begin{figure*}[htp]
   \centering
\parbox{7.5cm}{
\resizebox{7.5cm}{!}{\includegraphics{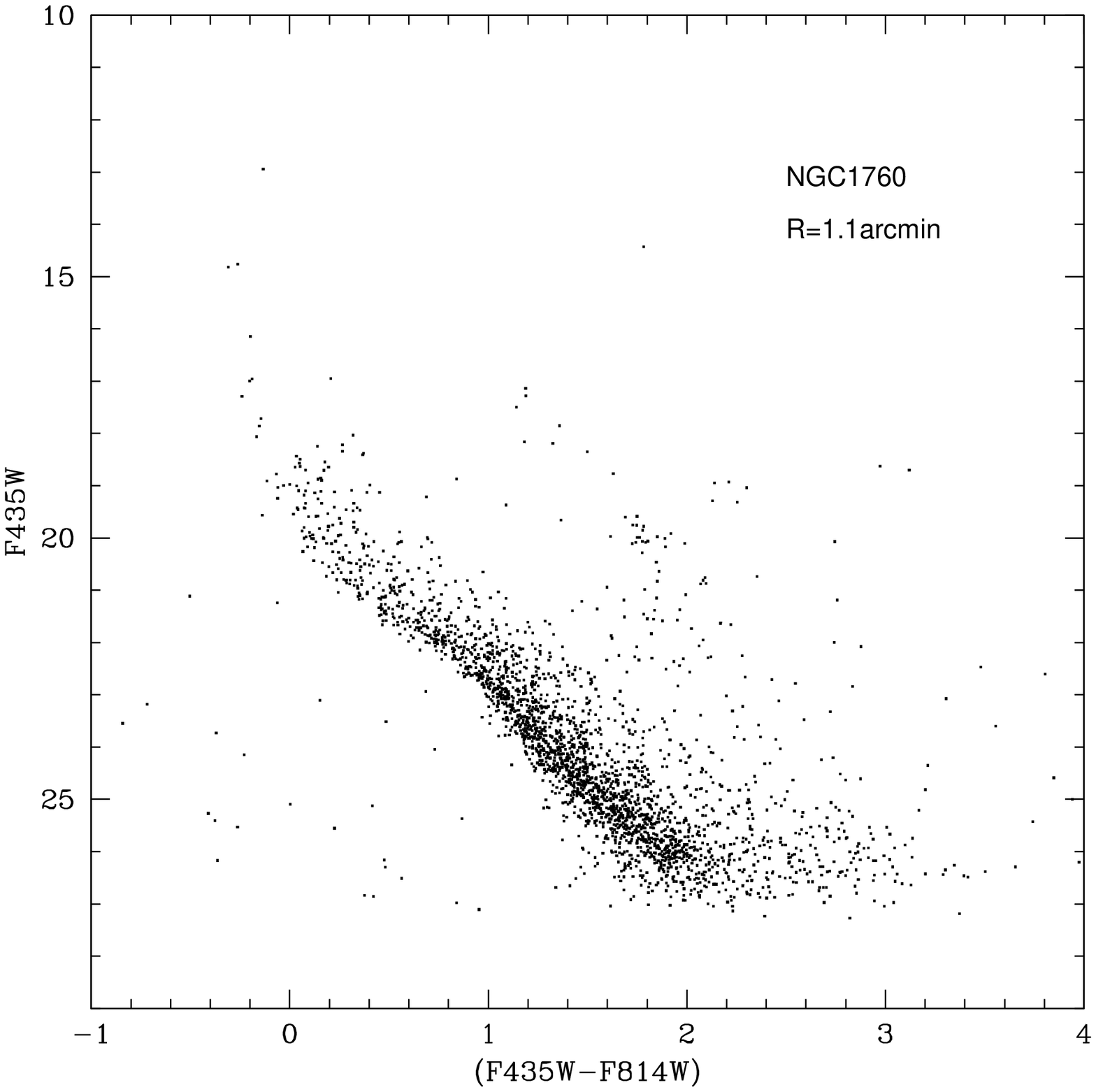}}\\
}
\parbox{7.5cm}{
\resizebox{7.5cm}{!}{\includegraphics{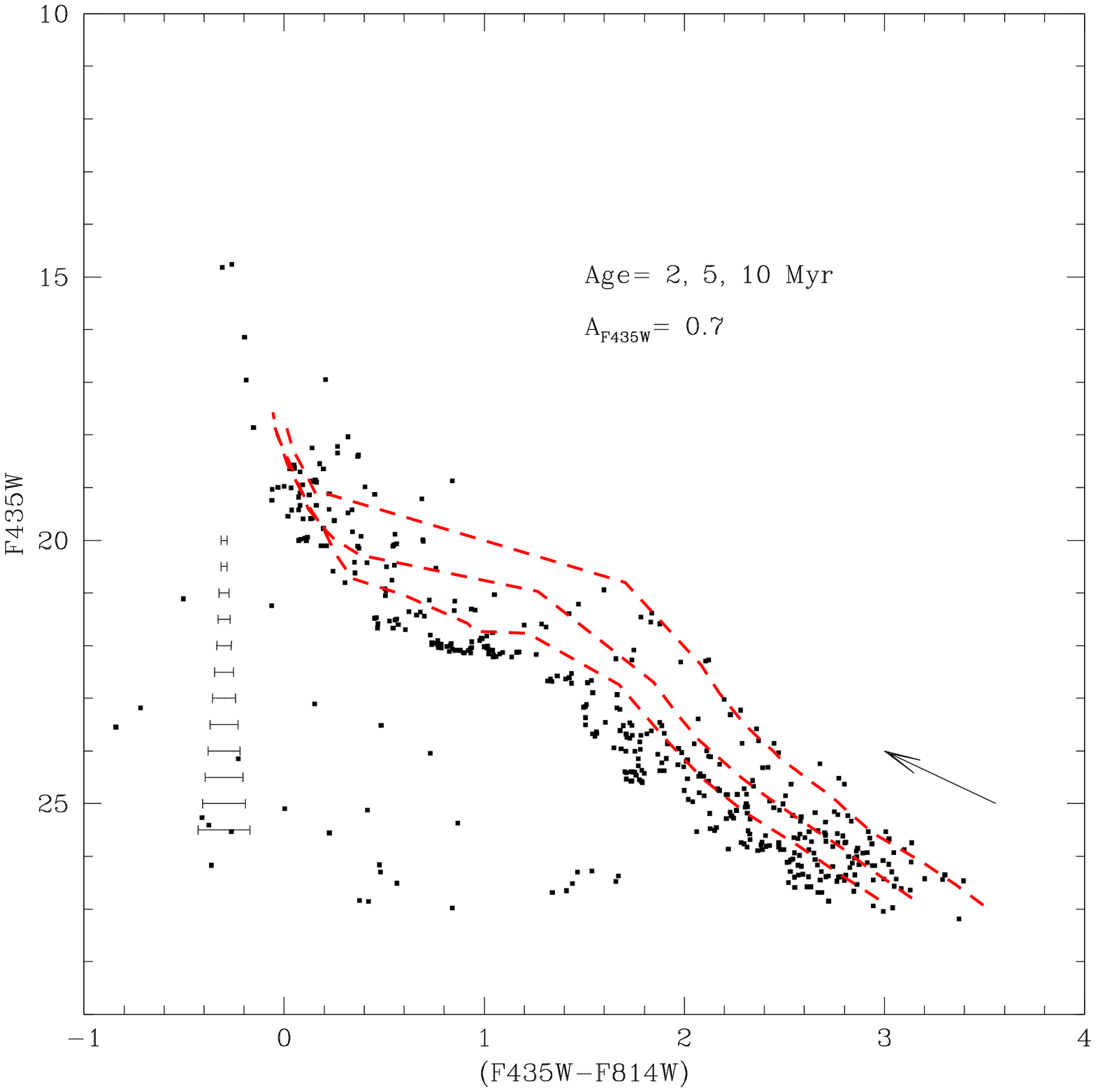}}\\
}

\parbox{7.5cm}{
\resizebox{7.5cm}{!}{\includegraphics{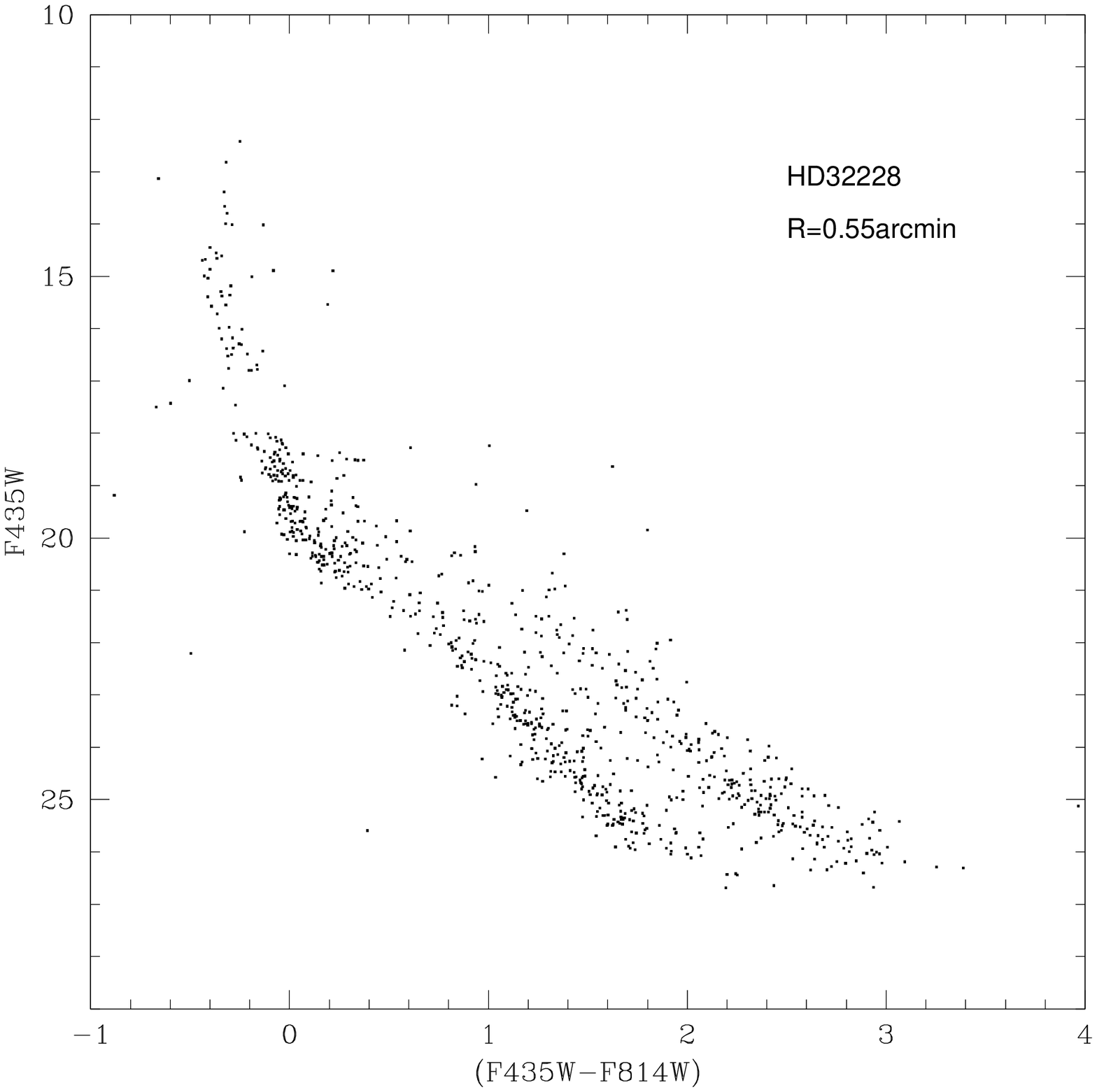}}\\
}
\parbox{7.5cm}{
\resizebox{7.5cm}{!}{\includegraphics{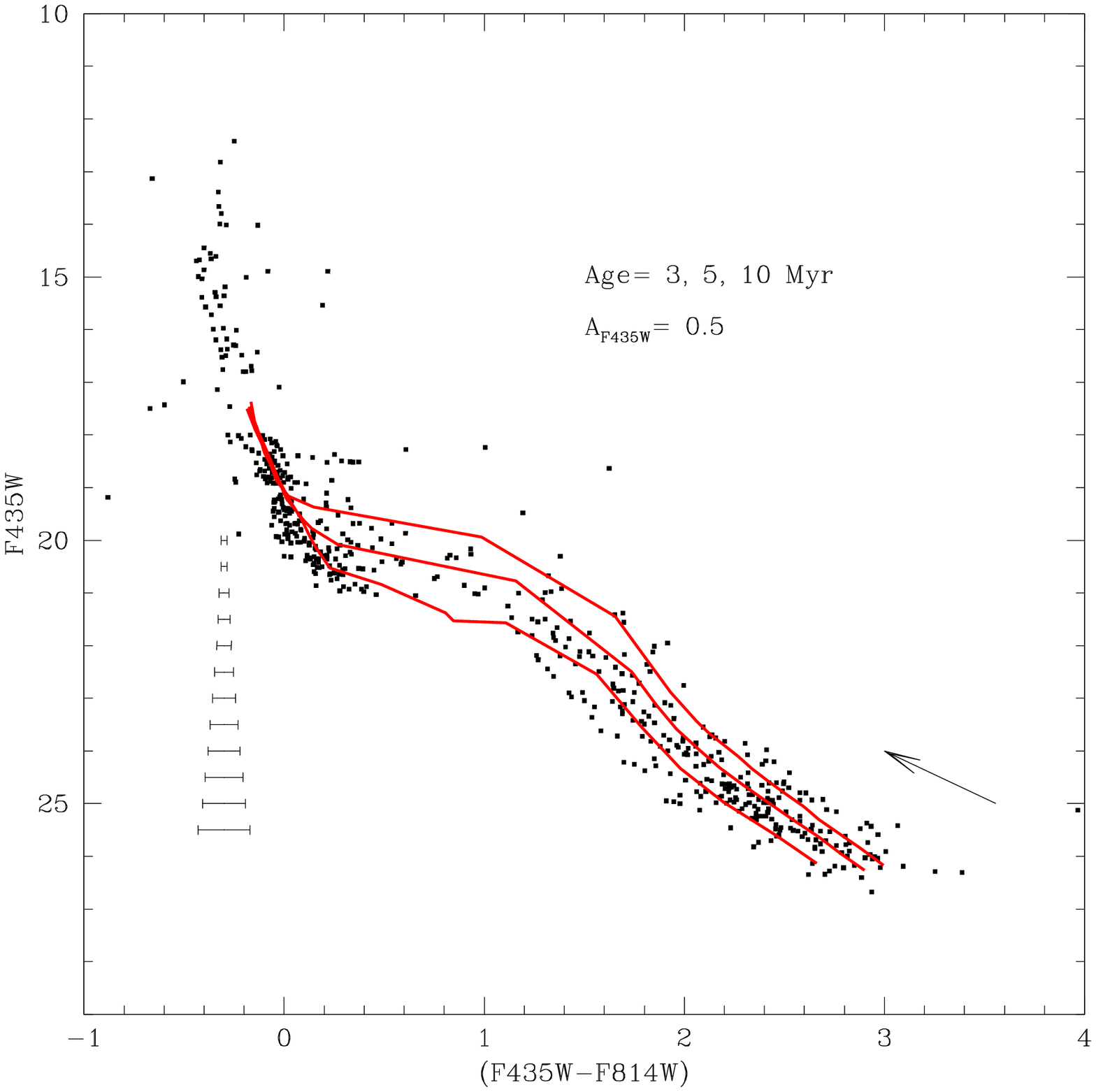}}\\
}
 \caption{CMDs of NGC~1760, HD~32228 in LH~9 (left panels). CMDs of the same objects when the field population is subtracted (right panels). The bars show the photometric errors on the color. PMS isochrones by Siess et al. (2000) are superimposed on the data. The arrow indicates the extinction vector.}
\label{cmd_2.fig}
\end{figure*}

\section{Results: Ages and stellar content of the associations inside N~11}
\label{ind_cmd}

 Here we analyze the CMDs of the most relevant  associations inside N~11 with the aim of deriving the age of the stellar population.
We assume a mean value of Z=0.19 for the metal content of the association, in agreement with \citet{pagel1999}, \citet{2000A&A...360..133D}.\\

\begin{figure*}[htp]
   \centering
\parbox{8cm}{
\resizebox{8cm}{!}{\includegraphics{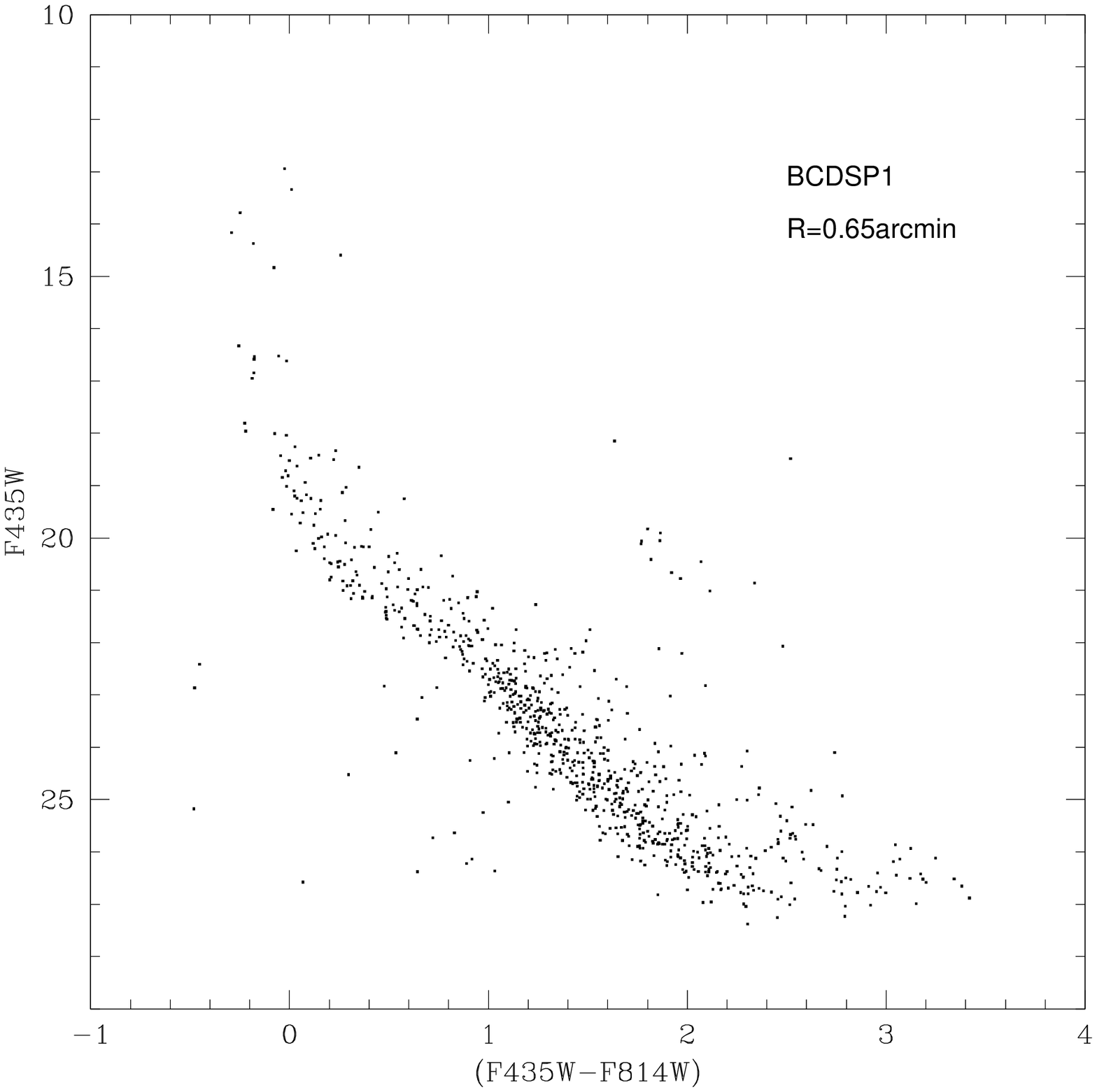}}\\
}
\parbox{8cm}{
\resizebox{8cm}{!}{\includegraphics{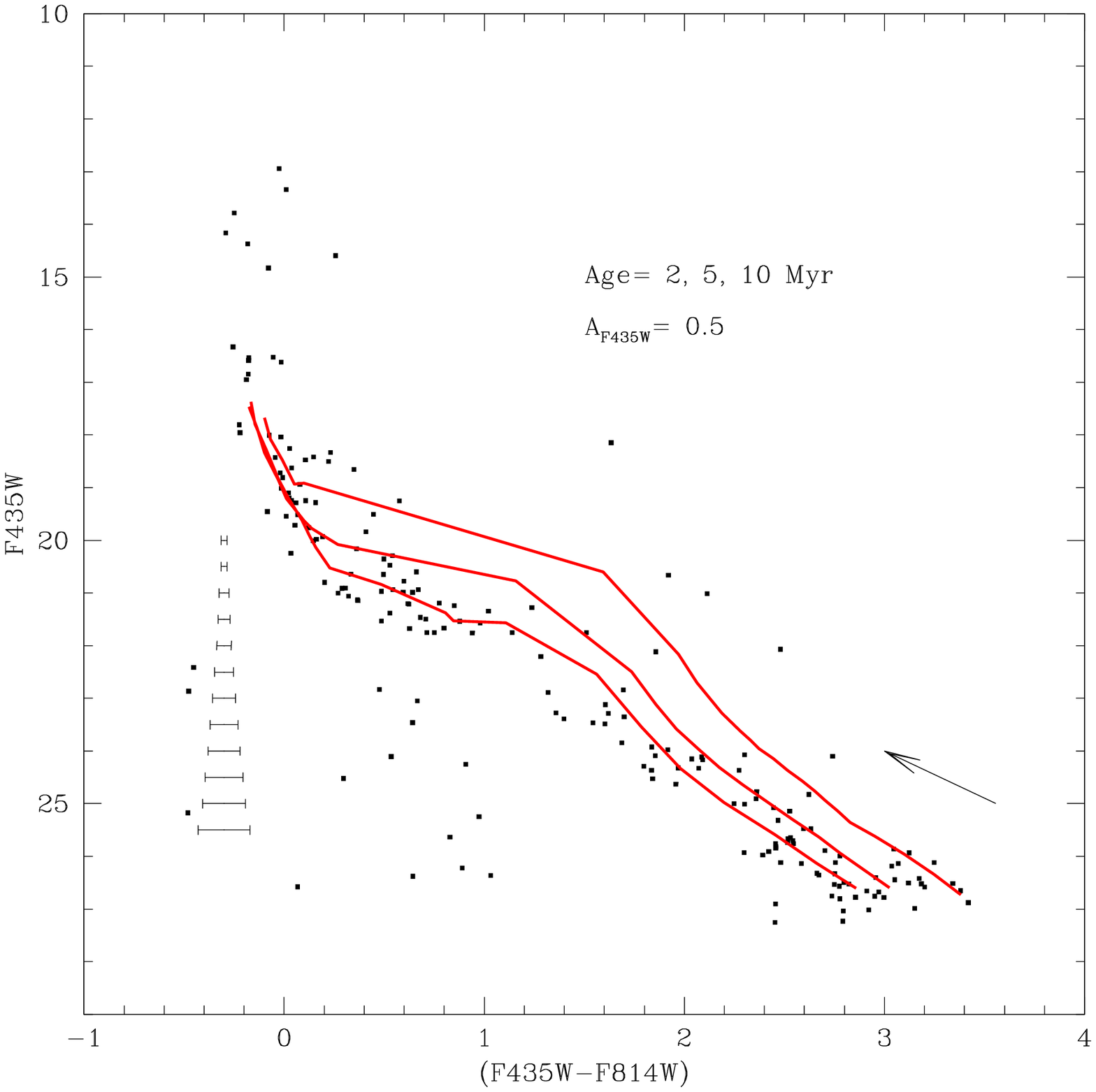}}\\
}

\parbox{8cm}{
\resizebox{8cm}{!}{\includegraphics{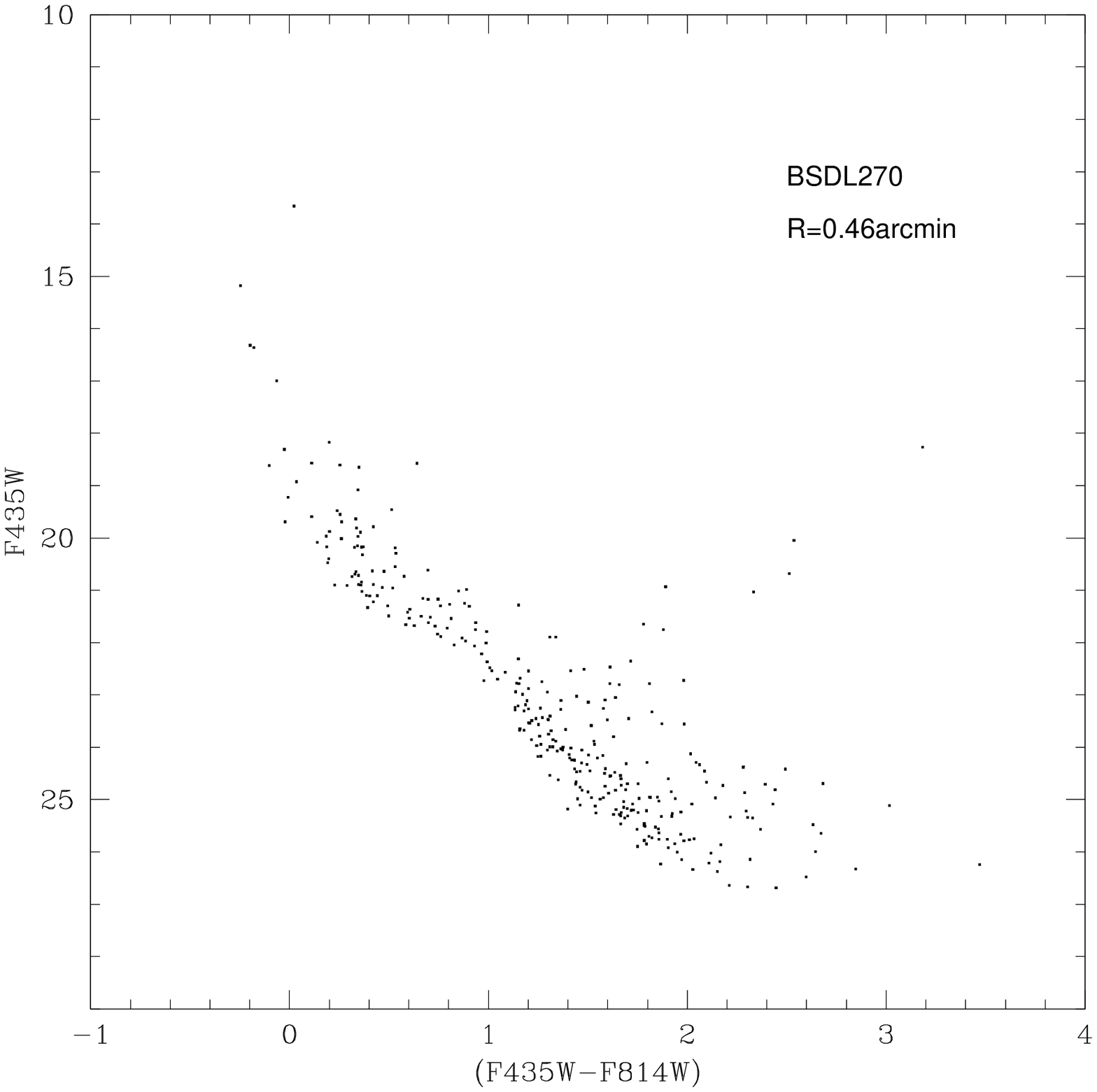}}\\
}
\parbox{8cm}{
\resizebox{8cm}{!}{\includegraphics{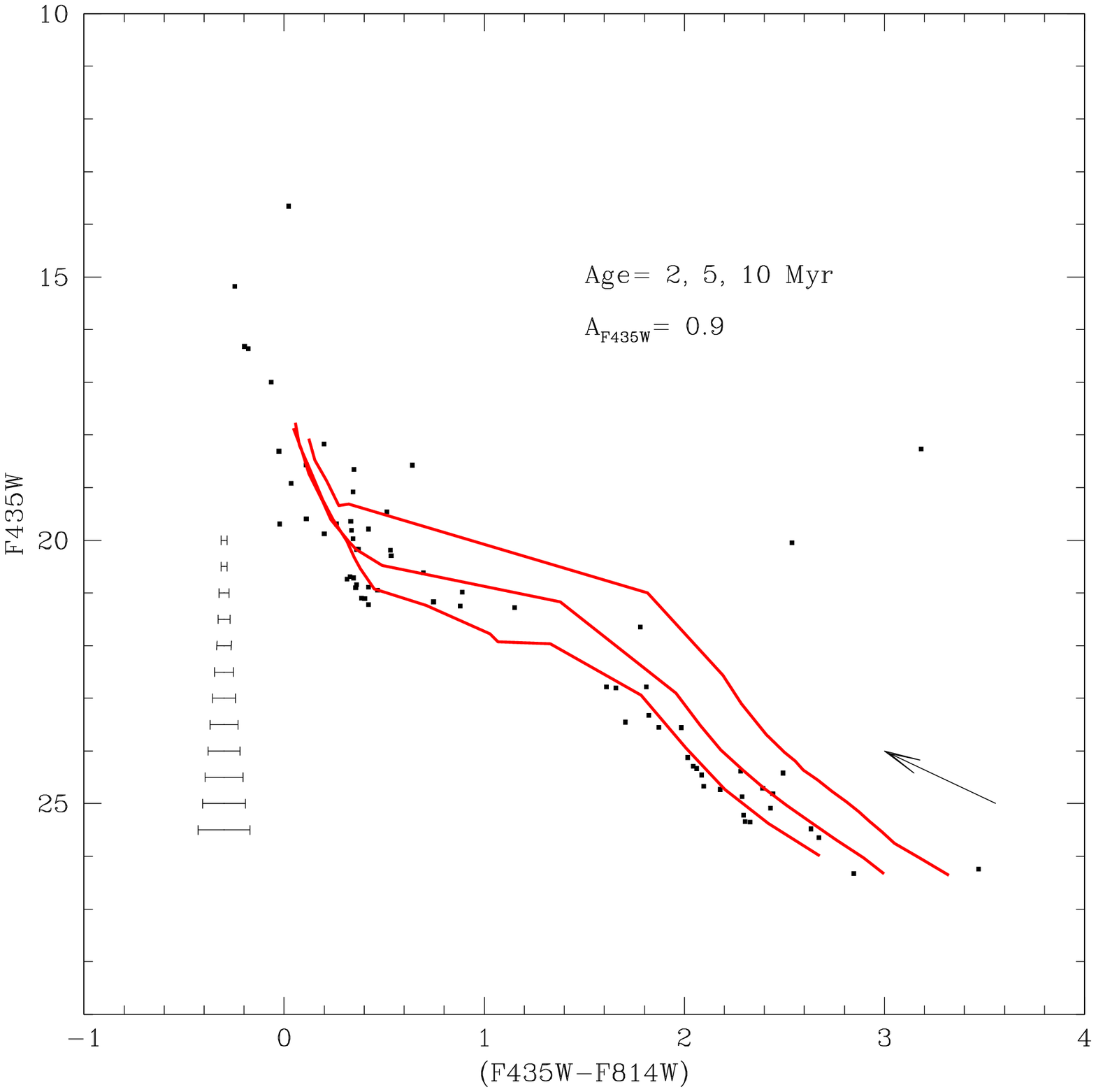}}\\
}

 \caption{CMDs of BCDSP~1, BSDL~270 in LH~9 (left panels). CMDs of the same objects when the field population is subtracted (right panels). The bars show the photometric errors on the color. PMS isochrones by Siess et al. (2000) are superimposed on the data.The arrow indicates the extinction vector.}
\label{cmd_3b.fig}
\end{figure*}

\begin{figure*}[t!]
\centering
\parbox{8cm}{
\resizebox{8cm}{!}{\includegraphics{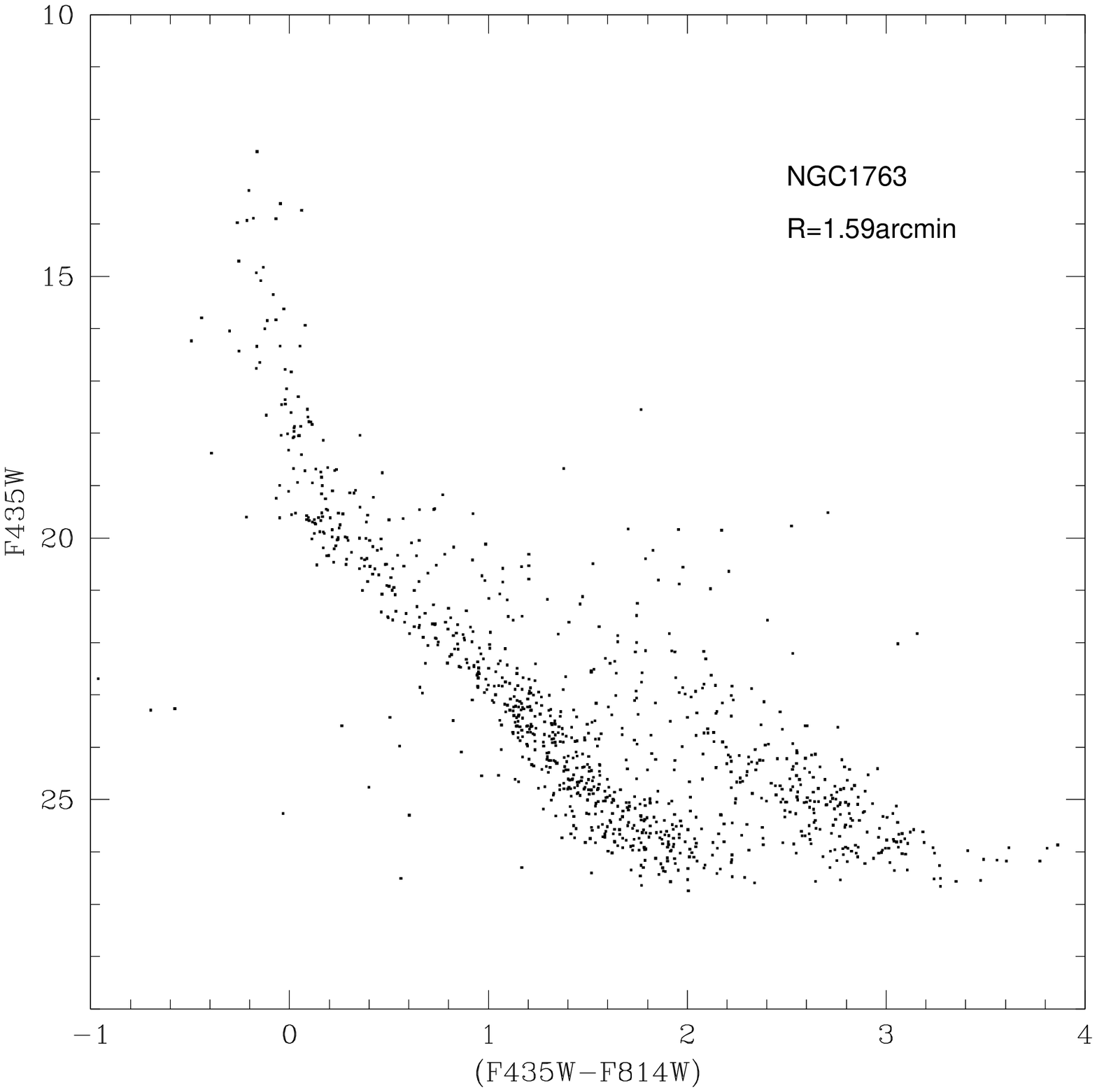}}\\
}
\parbox{8cm}{
\resizebox{8cm}{!}{\includegraphics{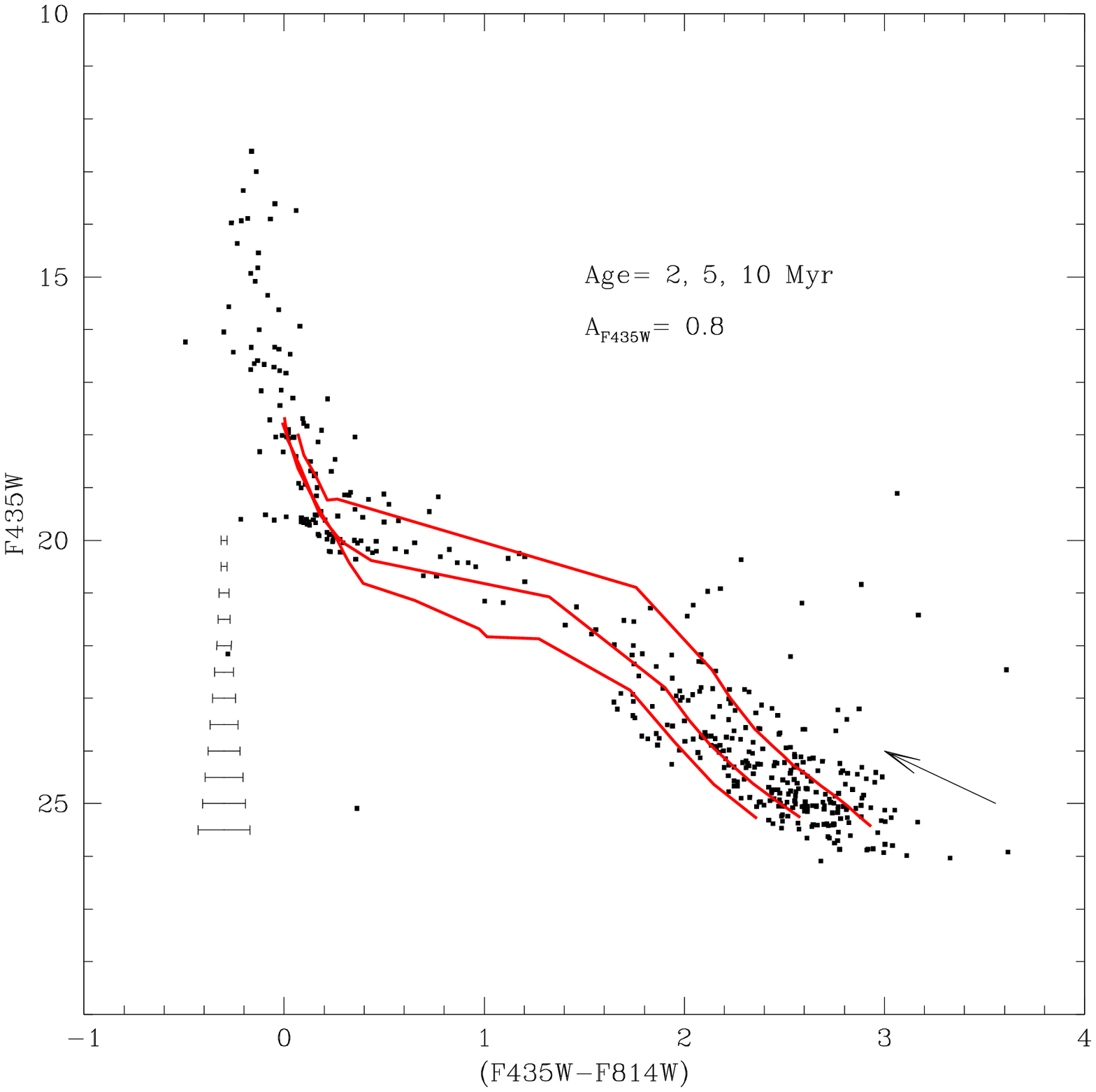}}\\
}
 \caption{CMD of NGC~1763 (LH~10) (left panel). CMDs of the same objects when the field population is subtracted (right panel). The bars show the photometric errors on the color. PMS isochrones by Siess et al. (2000) are superimposed on the data.The arrow indicates the extinction vector.}
\label{cmd_3.fig}
\end{figure*}

Figure~\ref{cmd_1.fig}  present the CMD of the N11 region covered by the HST/ACS photometry. A well-defined upper main sequence  of young blue stars is present.
Below the turnoff, moving to fainter magnitudes, the main sequence in
the CMDs of the associations and of the field population  is
increasingly densely populated. A striking difference at faint magnitudes between  field and association
CMDs is evident in the region redder than the MS, where a well-populated sequence  almost parallel to the MS itself
is found in the associations, but is completely absent in the field region (see Figs.\ref{sfr_field} and  \ref{cmd_1.fig}).
These stars  fit the location of the PMSs previously discovered in other MC associations very well \citep{2007ApJ...665L..27G, 2006ApJ...636L.133G}.\\
Suspicion might arise that the stars we identify as PMSs are instead highly reddened MS objects.
However, the fact that the interstellar extinction vector in the passbands F814W-F435W is  almost parallel to the PMS sequence(see Figs. \ref{cmd_2.fig}, \ref{cmd_3b.fig}, \ref{cmd_3.fig}) means that the interstellar extinction has only a small effect on the age of the PMSs.

\subsection{The stellar population in LH~9}

The most relevant associations/clusters in LH~9 in the region observed by HST are NGC~1760, NGC~1761, HD~32228, BCDSP~1, BSDL~270.

Figures~\ref{cmd_2.fig} and \ref{cmd_3b.fig} present the CMDs of these  objects  before and after the field star subtraction.
The PMS population is clearly visible in  all the fields. 
 The  CMDs after the field star subtraction are compared with pre-main sequence isochrones by \cite{2000A&A...358..593S} converted from the Johnson-Cousins pass-bands to the HST/ACS pass-bands using the transformations by \cite{sirianni2005}. Below we discuss the age determination of each object. {As a general remark, we remind  that it is difficult to derive a precise age determination of the PMSs on the basis of the CMDs only. Several effects such as photometric errors, binarity, differential extinction across the field and age spread in the population cannot be disentangled \citep[see for a detailed discussion] []{2008ASPC..384..200H}.  This holds for all the objects studied here. This problem will be discussed in the following sections.}

\begin{figure}[]
\parbox{8.5truecm}{
\resizebox{8.5truecm}{!}{\includegraphics[width=8truecm,height=5truecm]{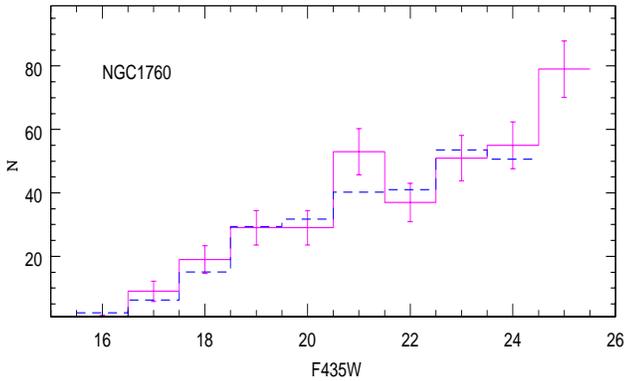}}\\
}
\caption{Observational LF of NGC~1760 corrected for interstellar extinction, {   completeness}, and field star subtraction (solid line) is compared with theoretical LF for a prolonged star formation in the age range 2$-$10 Myr as derived  using Siess et al. (2000) isochrones (dashed line) for stars fainter than F435W $\sim 17$.  }
\label{ngc1760_lum.fig}
\end{figure}

{\bf  NGC~1760.}
 Comparing the location of the PMSs with isochrones, we derive a tentative estimate of the age  in the range of 2$-$10 Myr. (see Fig. \ref{cmd_2.fig}).  More stringent constraints can be derived on the basis of the luminosity function. The observational LF after incompleteness correction, field star subtraction and extinction correction is compared with the theoretical one expected from the  \cite{2000A&A...358..593S} PMS isochrones. We make use of an authomatic $\chi^2$ procedure to find the best fitting solution.
{   The LF of the PMSs (see Fig.~\ref{ngc1760_lum.fig}) is suggesting  a prolonged formation from  2 to 10 Myr (for single stars)  with a IMF  slope $\Gamma =-3.0 \pm 0.3$ (when the Salpeter value is  $\Gamma =-2.35$). In the best-fit solution, the majority of the stars (about 75\% of the objects brighter than F435W $\sim$ 24.5) are however older than 6 Myr, while the youngest population (2 Myr) can account for less than 1\% of the population.
 The mass range of the PMSs is from 1.5 to 3.5 M$_\odot$ and 1.38 to 1.7 M$_\odot$  for 2 and 10 Myr, respectively. The lower mass limit corresponds to F435W $\sim 24.0$. Our value of the IMF slope agrees with \citet{1992AJ....103.1205P} who find  $\Gamma= -2.6 \pm 0.1 $  for masses higher than 8$-$9 M$_\odot$.

 The question at hand is  how reliable the age and age spreads are which we find  from the CMD and LF analysis.
With regard to the age spread,  photometric errors and binarity can  complicate the analysis. Binarity can be distinguished from age spread only when empirical luminosities are known with high accuracy \citep{2009IAUS..258...81H}. For single stars, taking into account photometric errors, we can say that the percentage of stars older than 6 Myrs is ranging from 65\% to 83\%. On the basis of this discussion, it  would be unrealistic to reach a more precise conclusion concerning the age distribution. Multiplicity surveys found a relatively high binary frequency ( $\ge$60\%) for masses higher than solar in young associations in the Galaxy  \citep[among others][]{1991A&A...248..485D}.    If we impose that
60\% of stars are in binary systems, we do not derive a significantly different age spread.

In addition to these effects, it is well known  that systematic trends are found in the literature using  various sets of pre-main sequence tracks. This holds in particular
 at sub-solar masses, while at solar masses the agreement is better.
In addition, an age-with-mass trend is found, i.e.  higher stellar masses are predicted to be older than lower mass stars in the same objects using the same set of stellar tracks. All this can easily account for 20\% to 100\% of the age difference when comparing the ages we derive with values taken from various literature sources \citep[see among others][]{2008ASPC..384..200H, 2009IAUS..258...81H}.  Taking these uncertainties into account,  we find a remarkably good agreement of our PMS ages with the literature spectroscopic determination of the ages of blue stars. }
OB stars are found in NGC~1760 by \cite{2007A&A...465.1003M}. This  sets a limit to the age at 7.0 $\pm 1 $ Myr. These authors find that the age range is broad in the whole association, going from about 7 to 1$-$2 Myr.

\begin{figure}[]
\parbox{8.5truecm}{
\resizebox{8.5truecm}{!}{\includegraphics[width=8truecm,height=5truecm]{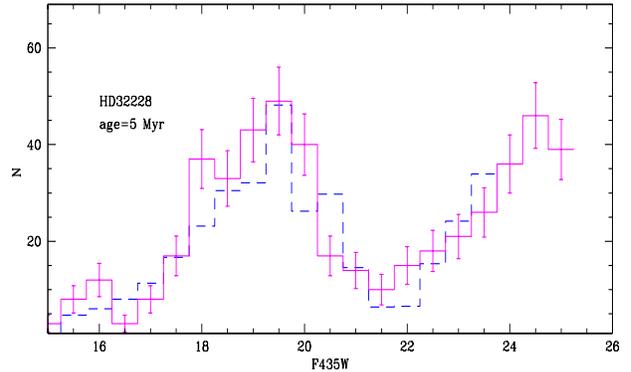}}\\
}
\caption{Observational LF of HD~32228  (solid line) is compared with theoretical LF for single stars derived  using Siess et al. (2000) isochrones (dashed line) for stars fainter than F435W $\sim 17$. (see text for details) }
\label{hd32228_lum.fig}
\end{figure}

{\bf HD~32228.}
A high  concentration of PMSs is found in HD~32228 located at the periphery of the association LH~9. {   Comparing the location of the PMSs with isochrones  we find that the age of the PMS is  in the range of 3$-$10 Myr (see Fig. \ref{cmd_2.fig}). If we assume that no binary stars are present, the LF of HD~32228 PMS after field star subtraction and extinction correction suggests  that the dominant population has an age of 5 $\pm 1$ Myr and a IMF slope   $\Gamma= -2.0 \pm 0.2 $ (see Fig.~\ref{hd32228_lum.fig}).  The mass range of the PMSs reaches from 1.5 to 2.5 M$_\odot$ when F435W $< 24.0$. However, a small age spread cannot be ruled out: solutions including, in addition to the main population of 5 Myr, about 25\% of the stars are as old as 7 Myr, and a small fraction   younger than 5 Myr cannot be ruled out.  Assuming that about 60\% of the stars are in binaries, we derive a slightly older age of the dominant population (6 $\pm$ 1 Myr), with a Salpeter IMF.}
The age of the cluster HD~32228  derived from spectroscopic of bright stars is of  about 3$-$4 Myr  \citep{1992ApJ...399L..87W} based on the presence of a WC star.

{\bf  BSDL~270  and BCDSP~1.}
 A small population of PMSs is found in these two small objects located at the periphery of LH~9. Comparing the location of the PMSs with isochrones in  BSDL~270 and BCDSP~1,  we find that a reasonable estimate of the age is  in the range of 5$-$10 Myr (see Fig. \ref{cmd_3b.fig}).
Due to the scarcity of the PMS number, the LF cannot reliably give information about the age of the stars.  Several B stars are found in these clusters, and one O6.5V (N11-065)  is in BCDSP1 \citep{2007A&A...465.1003M}.  N11-065 has a location on the HRD compatible with an age of about 1~Myr. However, \citet{2007A&A...465.1003M} find that this object has high He abundance. Using the appropriate isochrones for the surface He abundance, they derive an age of 6 Myr.
It is quite remarkable that in the whole region inside the uncertainties there is a good agreement between the ages of young blue  stars and of the red PMSs.

\subsection{The stellar content of LH~10}

\begin{figure}[b!]
\parbox{8.5truecm}{
\resizebox{8.5truecm}{!}{\includegraphics[width=8truecm,height=5truecm]{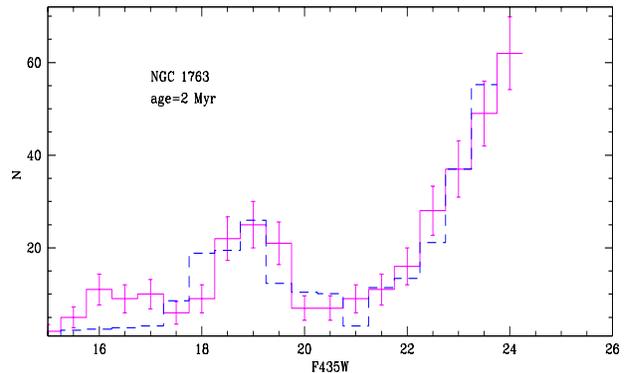}}\\
}
 \caption{Observational LF of NGC~1763  (solid line) is compared with theoretical LF (single stars) derived  using Siess et al.(2000) isochrones (dashed line) for stars fainter than F435W$\sim$17. }
\label{NGC1763_lum.fig}
\end{figure}

 The CMD of NGC~1763 in LH~10 before and after the field population is subtracted is shown in Fig.~\ref{cmd_3.fig}. The CMDs refers to a small region near the center of a radius $R\sim 1.6^\prime$.  PMSs are clearly identified in the CMD. {   The LF for single stars is presented in Fig.~\ref{NGC1763_lum.fig} after incompleteness correction, field star subtraction and extinction correction.}
{   The comparison of the observational models with the theoretical one calculated using \cite{2000A&A...358..593S} models suggests that the dominant population has an age of 2$-$3  Myr, using a IMF slope $\Gamma = -2.20 \pm 0.2$. The  PMS mass range  is 1.5$-$3.5 M$_\odot$. As we discussed in the previous section, it is difficult to completely discard solutions having a small percentage of older objects.} This result agrees with \citet{1992AJ....103.1205P} who find that the slope of the initial mass function of the young blue MS stars in LH~10 is $\Gamma =-2.1 \pm 0.1$, significantly flatter than that of LH~9. {   If we account for 60\% of binaries, we derive an age of 3 $\pm 1$ Myr, with a steeper IMF ($\Gamma = -2.80 \pm 0.2$).}
Using bright blue star spectroscopy, \cite{2000A&A...361..877H} estimate that LH~10 in N11B is 3$\pm 1$ Myr old. \cite{2007A&A...465.1003M} derive a broad age spread in LH~10 from 1 to 6 Myr. But the majority of the stars are younger than 4.5 Myr, and only one object (N11-087) in the cluster core is older than this age. These authors conclude that the mean age is 3$\pm 1$ Myr.
 Herbig Ae/Be stars are found inside the association, together with  a methanol maser \citep{2006AJ....132.2653H}.  


\begin{figure*}[t!]
\centering
\parbox{8cm}{
\resizebox{8cm}{!}{\includegraphics{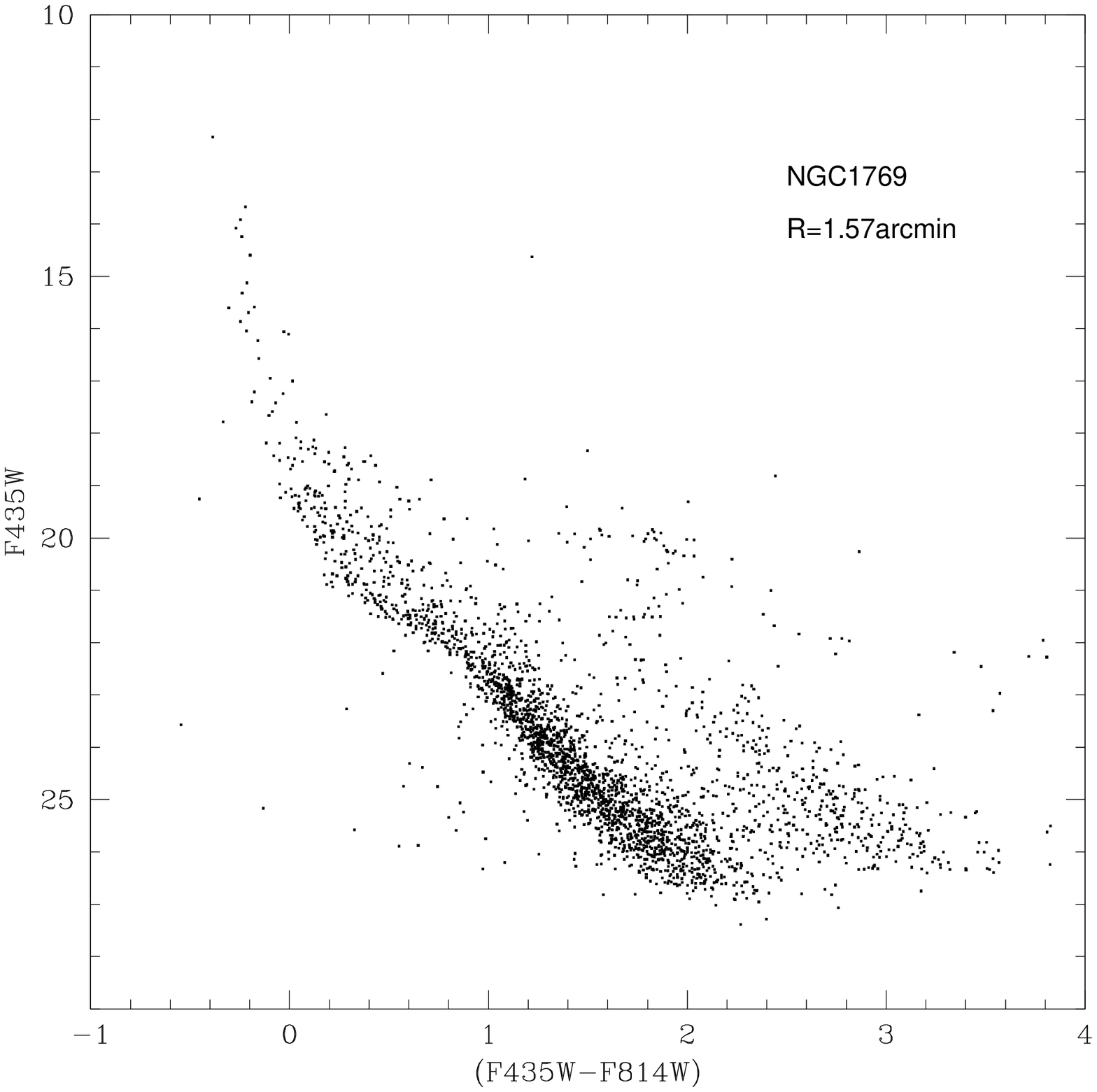}}\\
}
\parbox{8cm}{
\resizebox{8cm}{!}{\includegraphics{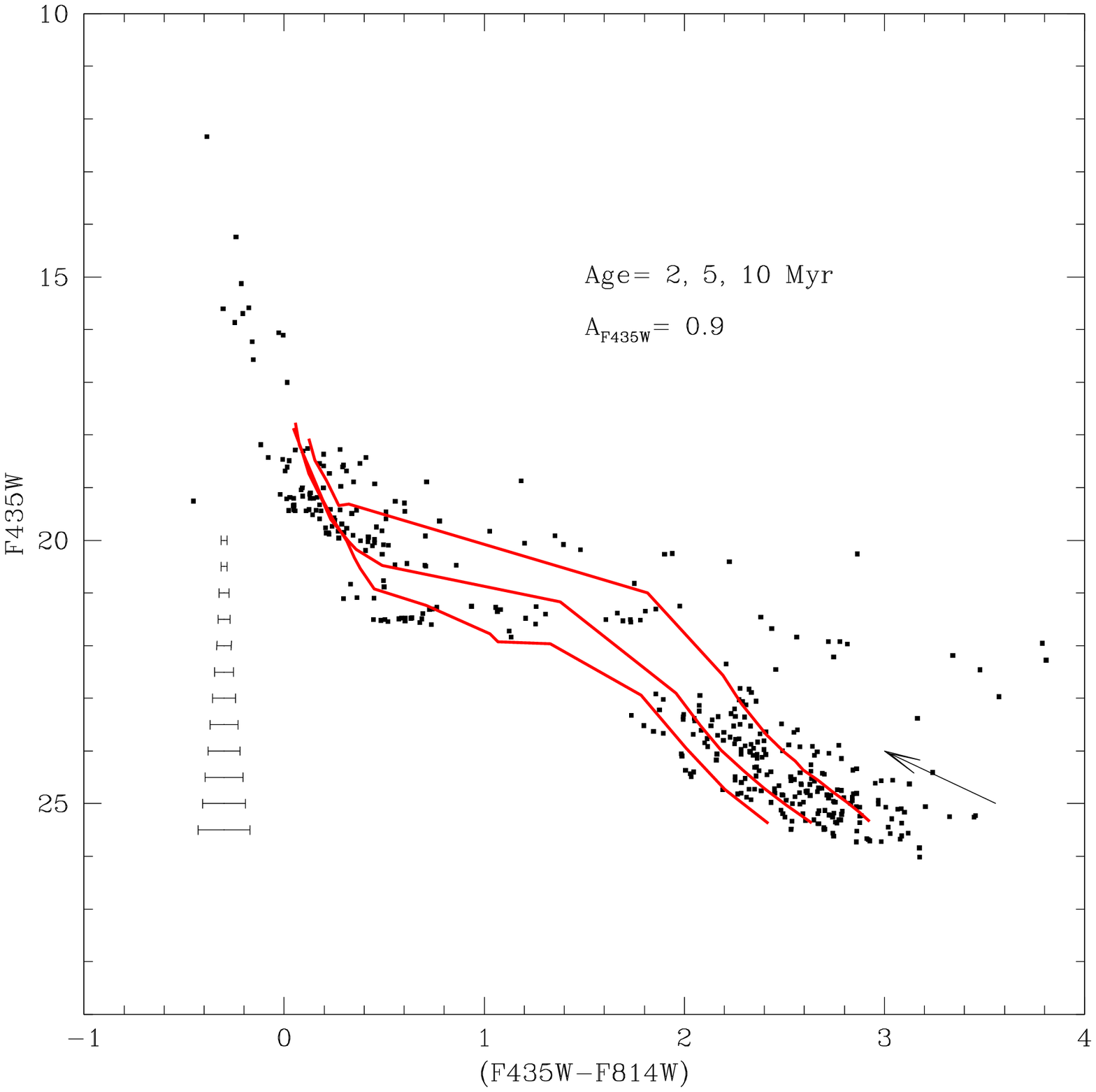}}\\
}
\\
\parbox{8cm}{
\resizebox{8cm}{!}{\includegraphics{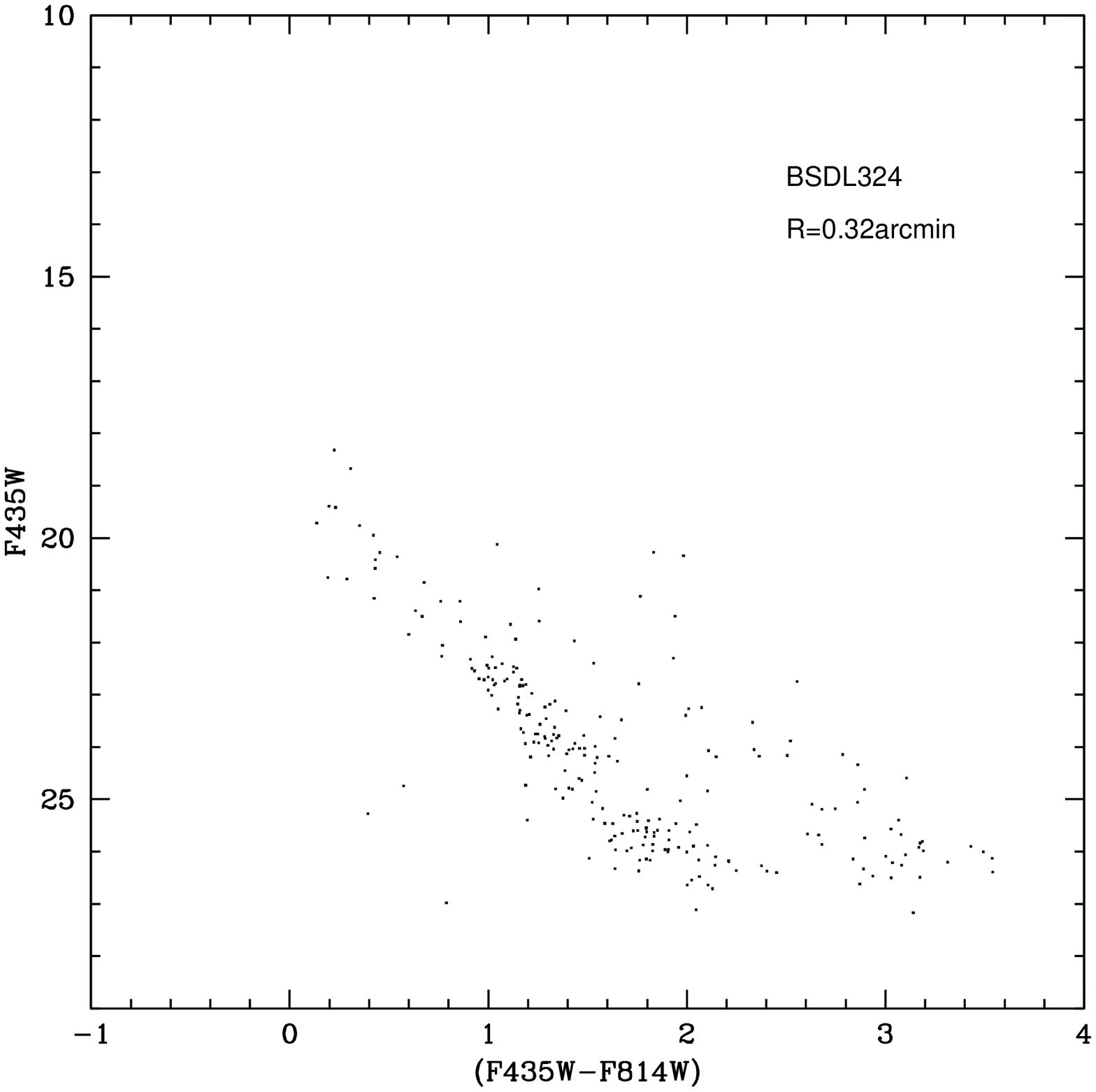}}\\
}
\parbox{8cm}{
\resizebox{8cm}{!}{\includegraphics{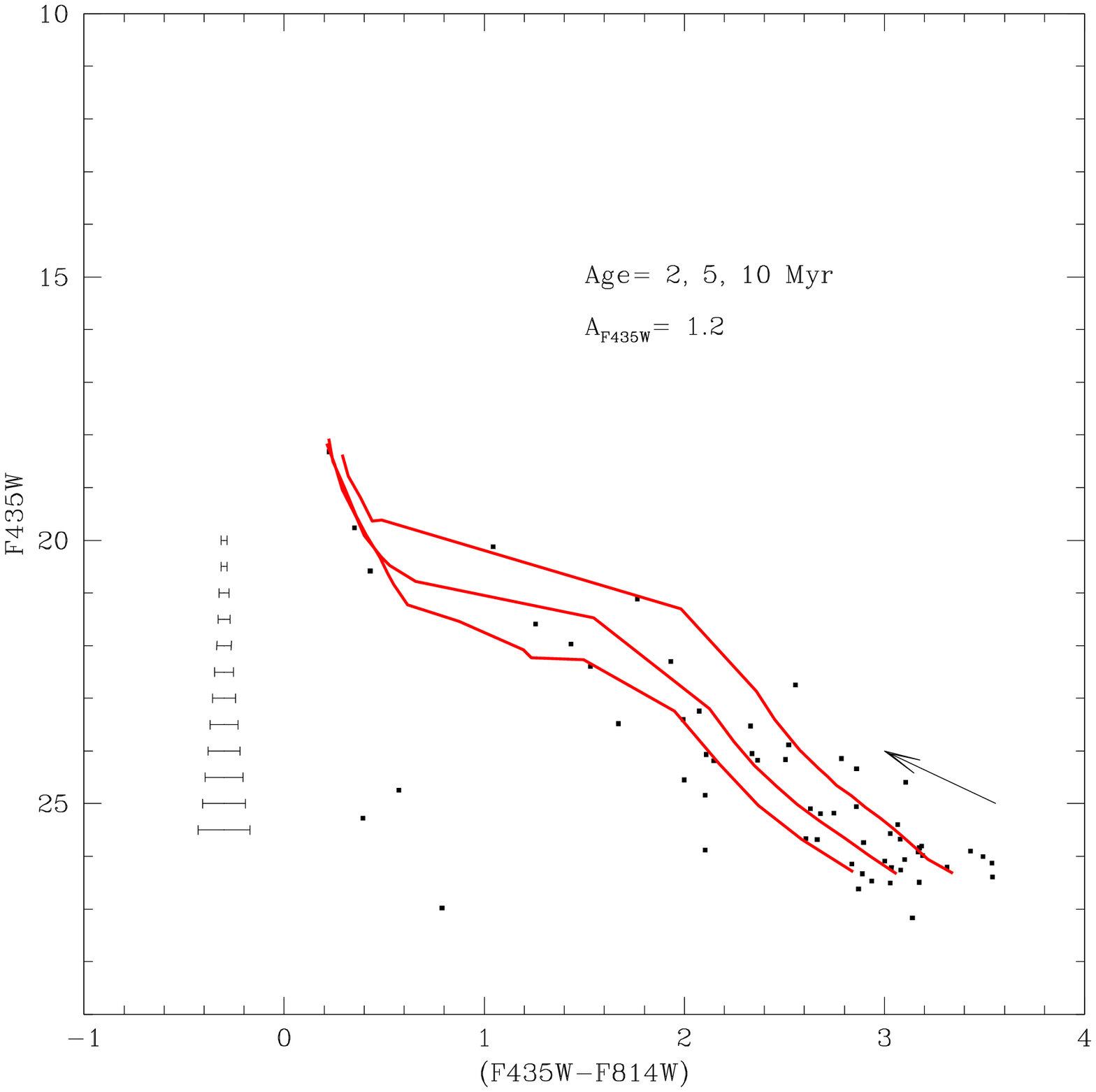}}\\
}
 \caption{CMDs of NGC~1769 and BSDL~324 (in LH~13) (left panels). CMDs of the same objects when the field population is subtracted (right panels). The bars show the photometric errors on the color. PMS isochrones by Siess et al. (2000) are superimposed on the data.}
\label{cmd_4.fig}
\end{figure*}

\subsection{The stellar population in LH~13}

\begin{figure}[!htp]
\parbox{8cm}{
\resizebox{8cm}{!}{\includegraphics[width=8truecm,height=5truecm]{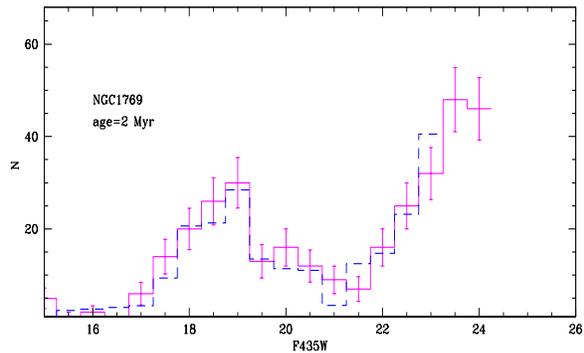}}\\

}

 \caption{Observational LF of NGC~1769 (solid line) is compared with theoretical LF derived  using Siess et al. (2000) isochrones (dashed line) for stars fainter than F435W $\sim 17$. }
\label{ngc1769_lum.fig}
\end{figure}

Our HST data in LH~13 include  NGC~1769 and  BSDL~324.  The CMDs of the association before and after the field population is removed are shown in Fig.~\ref{cmd_4.fig}. The comparison of the location  of PMS on the CMD  of  NGC~1769 with isochrones suggests  an age in the range of 2$-$5 Myr. {   Analyzing the PMS LF (see Fig.~\ref{ngc1769_lum.fig}) we find that the dominant population (about 85\% of the stars) have an age of 2$-$3  Myr with a Salpeter IMF when only single stars are considered. The PMS mass range  is 1.5$-$3.5 M$_\odot$.
Including $60\%$ of binaries, we derive an age of 3$-$4 Myr, with a  steep IMF slope  $\Gamma= -2.8 \pm 0.1 $.}
No information about the age of the PMS in BSDL~324 can be derived from the CMD, due to the small number of stars detected.
Spectroscopic determination of the age of OB stars  in LH~13 are in the range of 3$-$5 Myr \citep{2000A&A...361..877H}. A small concentration of  Herbig Ae/Be stars is found here \citep{2006AJ....132.2653H}.

\subsection{YSOs from near-IR archive data: data selection and object distribution}
\label{YSO}

\begin{figure}[h]
\centering
\resizebox{7cm}{!}{\includegraphics{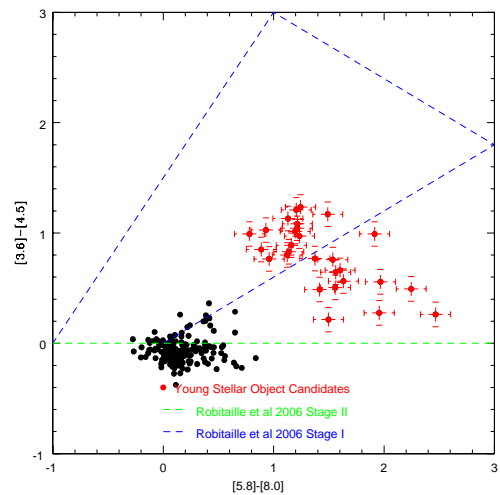}}
\caption{[5.8][8.0])-([3.6][4.5] color-color plot of the N~11 field.
Red identifies YSO candidates. Error bars show the nominal errors on the photometry as given by the SAGE catalog. The blue line outlines the region where Stage 0-I objects are found on the basis of \cite{2006ApJS..167..256R} photometric models. In the region redder than [5.8][8.0] $\sim$0 between the green and the blue line  Stage II/Stage III objects can be detected.
  }
\label{red1}
\end{figure}

\begin{figure}
\centering
\resizebox{7cm}{!}{\includegraphics{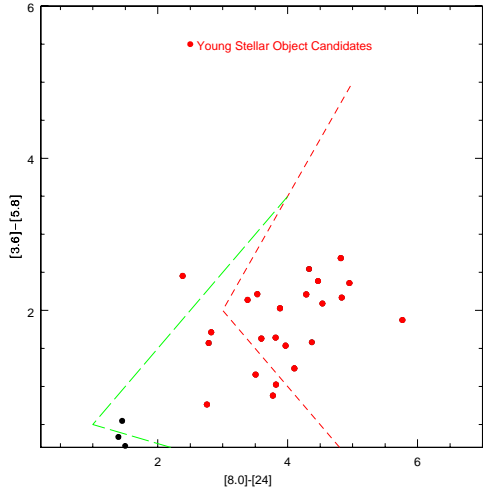}}
\caption{IRAC-MIPS color-color plot in the pass-bands ([8.0][24])-([3.6][5.8]). The short dashed red line separates the region where Stage 0-I objects are found, on the basis of \cite{2006ApJS..167..256R} photometric models. The long-dashed green line shows the region where Stage II objects can be detected.}
\label{red2}
\end{figure}

\begin{figure}
\centering
\resizebox{7cm}{!}{\includegraphics{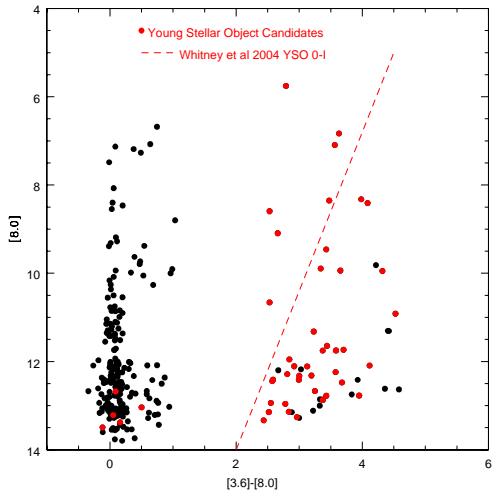}}
\caption{[8.0]-([3.6]-[8.0]) color-magnitude diagram compared with the  \cite{2004ApJ...617.1177W} models. The objects redder than the line are likely to be YSOs of the Stage 0-I. }
\label{red3}
\end{figure}

In this section  the detection of YSOs on the basis of IR Spitzer archive  data is presented.
The comparison of the observational color-color plots in the infrared passbands with the models by \cite{2006ApJS..167..256R} allows us to select candidate  YSOs of Stage 0-I (presenting significant infalling envelope , and having ages of $<$ 0.1 Myr) and Stage II (showing a thick disk source, having 0.1 $<$age$<$ a few Myr) and finally, Stage III YSOs  sources (with an optically thin disk or no disk).  Here we adopt the classification by \cite{2006ApJS..167..256R}, which  is analogous to the traditional Class scheme \citep{1987IAUS..115....1L}.
 On the basis of 200,000 YSO models  computed at ten viewing angles, \cite{2006ApJS..167..256R} point out that the IRAC color-color plot ([5.8][8.0])-([3.6][4.5]) contains a large region where the fraction of  Stage 0-I/all YSOs is close to  100\%, a small intermediate region where  the fraction of Stage II/all  sources is higher than 80\%. 
IRAC-MIPS ([8.0][24])-([3.6][5.8]) are effective in separating stars with no circumstellar material from Stage 0-I and Stage II objects.
Here we restrict ourselves to study Stage 0-I and Stage II objects.
In Figs.\ref{red1} and \ref{red2} the SAGE archive data in the N~11 region are plotted in the IRAC color-color plot ([5.8][8.0])-([3.6][4.5]) and IRAC-MIPS color-color ([8.0][24])-([3.6][5.8], respectively to identify the  YSOs.
Finally YSOs candidates selected on the basis of previous plots are examined in the [8.0]-([3.6]-[8.0]) color-magnitude diagram and compared with the \cite{2004ApJ...617.1177W} models (see Fig.~\ref{red3}).
About 30 objects are identified as YSO candidates.
Background galaxies and foreground Milky Way stars are known to contaminate the IRAC and MIPS CMDs and color-color plot. In order to derive the expected contribution in the studied region, we select a control field located outside the star forming region, namely at $\alpha(J2000)$=4h 58m 40s, $\delta(J2000)$ -66$^0$ 34$\prime$ 47$^{\prime\prime}$, which has an area comparable to the area of the whole N~11 field.
Using the same diagnostic plots  we find six objects in the control field
redder than ([3.6]-[8.0])$\sim 1$ in the [8.0]-([3.6]-[8.0]) color-magnitude diagram  {and only 3 (over 30 in total) in the region occupied by YSO candidates } in the ([5.8][8.0])-([3.6][4.5]) color-color plot.  Only 10\% of the YSO candidates  having [8.0]$< 14$ are expected to be background objects. The percentage of background galaxies agrees with the statistics by \cite{2006AJ....132.2268M}.

The YSO spatial distribution is shown in Fig. \ref{ima1}: type I and II objects having ages from 0.1 to 1 Myr  are found at the same location as the PMSs, indicating that several generations of stars are present in the region. The youngest candidates (type I) are found in NGC~1760 (LH~9), north of BCDSP1 (LH~9) and NGC~1769 (LH~13).

\subsection{Toward a coherent picture: the PMS, YSOs, OBs, Herbig Ae/Be  and their age distribution}
\label{YSO_PMS}

\begin{figure}[h]
\centering
\parbox{9cm}{
\resizebox{9cm}{!}{\centering {\includegraphics{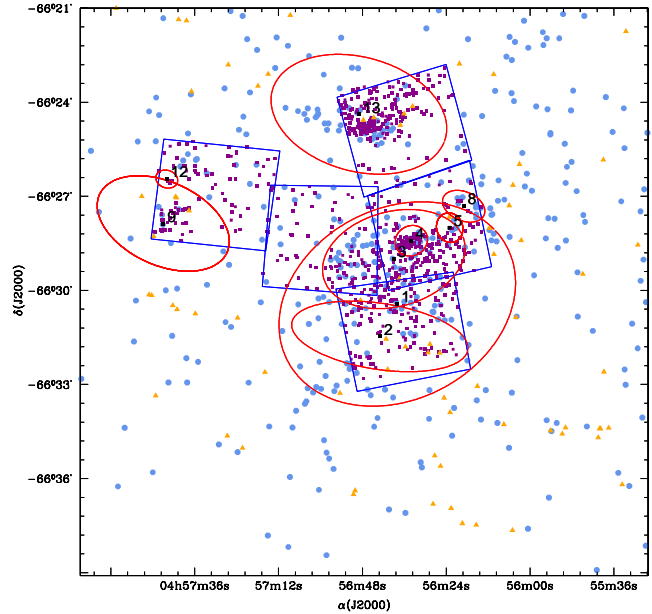}}}
}
 \caption{Bica clusters and associations  are overplotted on the distribution of PMS (dark magenta dots), the concentration of OB stars (light blue circles) and Herbig Ae/Be stars (orange triangles) (from Hatano et al 2006). Red ellipses represent Bica et al. object dimensions (see Fig.~1 for details). The blue contours show the observed area.}
\label{ima2}
\end{figure}

In Fig. \ref{ima2} we compare the spatial location of the PMSs  with the location of  Bica et al clusters and associations in the area and with the distribution of the OB stars and Herbig Ae/Be candidates \citep{2006AJ....132.2653H}.  
Summarising the results of the previous sections, the ages derived in literature from spectroscopy of blue stars and from PMS analysis are quite consistent in the whole area.
 The analysis of HST data implies that PMSs in LH~9  South the majority of the stars are  slightly older than the remaining part of N~11, while LH~9 North, LH~10(center) and LH~13 seem to be younger (with ages of  5, 2, 2 Myr).  A large age spread from 1 to 6 Myr is found from the spectroscopy of blue stars in these regions and is confirmed by our analysis.  The presence of YSOs  in LH~9 South, LH~13 (type I) and LH~9 (North), LH~10 (Type II)  suggests that the star formation in the region is still going on. The most conspicuous concentrations of PMSs correspond to the location of OB stars. Herbig Ae/Be stars are found in all these locations, confirming the presence of a population of ages in the range of 1$-$3 Myr. All the indicators present a coherent picture of the star formation in the region.
 \cite{2007A&A...465.1003M} estimate that the time needed for a supernova shock to cross the distance between LH~9 and LH~10 is of the order of 0.1 Myr. Since the lifetime of a massive star is of the order of 3 Myr, the age difference agrees with the idea that
LH~9 has triggered the  formation of a first star generation in LH~10. This cannot be excluded on the basis of our discussion, even if more generations of stars are found in the region.

We compare the location of PMSs  and YSOs with H$_\alpha$ emission (Mac Low et al 1998) and CO emissions
(Israel et al 2003) (see Fig. \ref{ima1}). We find that  PMSs and YSOs are located near the maxima of the  H$_\alpha$ emission and at the border of the CO clouds in N11B and N11C. PMSs are found at the border of the cavity surrounding LH~9.

\section{Results:  clustering of PMS stars}
\label{clustering}

In this section we discuss the clustering degree of the PMSs, first using the two-point correlation function and then the MST method.{   Only the stars selected as PMSs on the basis of the HST/ACS photometry are considered.}

 The correlation function for the PMSs shows a well defined peak, while in the case of the whole star sample it is considerably flatter, implying a weaker correlation (see Fig. \ref{autocross}). The maximum value of the correlation function of the PMSs is comparable to the peak values found in Galactic star forming regions such as Rosetta Molecular cloud \citep{2005AJ....130..721L} or M16 \citep{2007ApJ...666..321I} and reflects the strong degree of clustering of the population.

The full width half maximum of the peak  for PMSs is of $\sim 10 $ pc, of the order of the typical size of molecular clouds in the SMC and LMC which is going from about 10 to 100 pc \citep{2003A&A...401...99I}. This reflect the structure of the interstellar medium from which the stars formed. {   This result may however be affected by  a potential bias due to the fact that the ACS pointings are not covering the whole region, but are mainly centered on well-known star forming regions, where higher concentrations of young stars are expected.}

\begin{figure}
\centering
\resizebox{7cm}{!}{\includegraphics{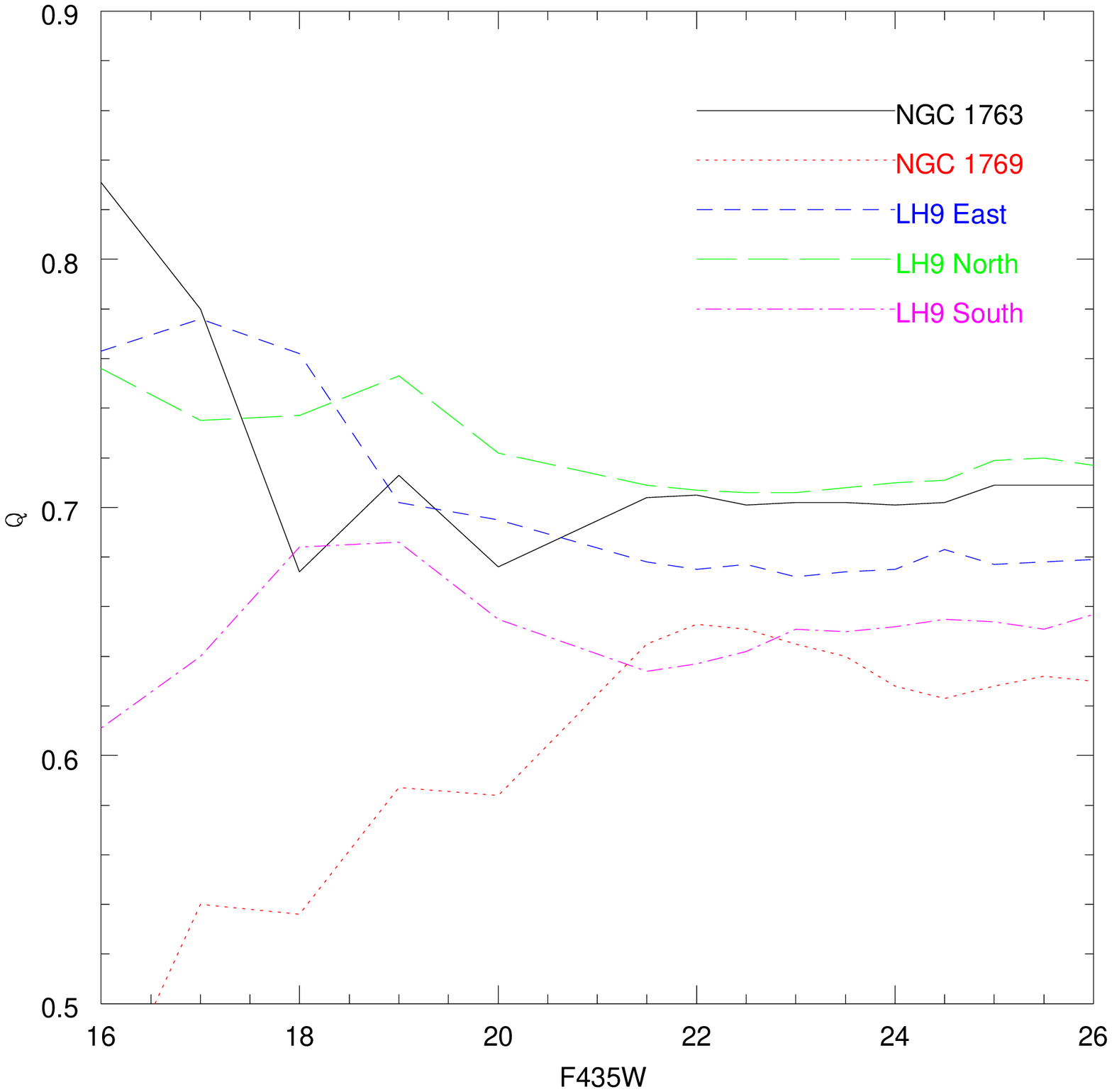}}
\caption{ \textsl{Q} values versus magnitude in the different fields is shown. Bright stars seem to be radially concentrated in the area of NGC~1763 and LH~9 North and East, where clusters (HD~32228) and concentrations of PMS stars and OB stars are located. An opposite trend is found in external areas like NGC~1769 and LH~9 South. }
\label{pmscum.fig}
\end{figure}

In Fig.~\ref{NGC_grafo.fig} we present the MST graphs of some of the clusters and associations in the region.
In Table \ref{Q_val2.tab}  we give the values of the \textsl{Q} parameter, calculated selecting the stars inside the elliptical areas described by the Bica et al. cluster parameters.
Although in principle \textsl{Q} is not expected to depend on the selected area,  \citet{2009MNRAS.392..868B} found a degeneracy in the sense that highly elongated structure like these show a lower \textsl{Q} value. The values in Table \ref{Q_val2.tab} include the ellipticity correction derived by \citet{2009MNRAS.392..868B}.   To avoid the uncertainties due to the shape, we calculate  \textsl{Q} using the quadrilateral areas which correspond to the observational CCD fields.%
 The results are presented in Table \ref{Q_val_ccd.tab} for all the stars. No significant changes are found due to the area selection.
To verify whether blue bright stars and fainter red PMSs have a similar clustering behavior, we  calculate   \textsl{Q} for PMSs fainter and brighter than F435W $\sim 20 $.  Tables \ref{Q_val_ccd_r.tab} and  \ref{Q_val_ccd_b.tab} respectively show the results. We use  quadrilateral areas.
Finally we derive  \textsl{Q} for the OB star candidates taken from the Hatano et al. catalog in an area of 0.24 sq. deg. in the centre of the N11 star forming region. The corresponding graph is presented in Fig. \ref{NGC_grafo.fig}.
The  \textsl{Q} parameter is derived for the whole area, without subdivision in smaller fields, due to the small number of OB candidates.
Figure~\ref{pmscum.fig} shows the trend of \textsl{Q} with the magnitude of the population.  \textsl{Q} is calculated for all the stars brighter than a given magnitude.
\noindent The main results can be summarized as follows:

\begin{itemize}

\item In all the fields related to N11, the values of \textsl{Q} $\leq$  0.8 are typical of clustered substructures.  Smaller objects have slightly scattered \textsl{Q} values in the range of 0.62$-$0.74.
For comparison, we remind that \citet{2004MNRAS.348..589C} find that
a value of \textsl{Q} $\sim 0.56, 0.66, 0.70$  corresponds to a projected fractal model with dimensions of 1.8,2.2, 2.4 respectively.

\begin{table}[h]
\caption{Values of Q for the all the stars inside the cluster areas given by the Bica et al parameters.}
\begin{center}
\begin{tabular}{ l c c c}
\hline
 Object  & $\bar{m}$   & $\bar{s}$ &  $\textsl{Q}$ \\
\hline

 NGC~1769 & 0.86    & 1.33  &0.65 \\
 NGC~1763 & 0.73    & 0.98  &0.75  \\
 NGC~1761 & 0.77    & 1.18  &0.66  \\
 NGC~1760 & 0.73    & 1.20  &0.56  \\
 HD~32228 & 0.73    & 0.93  &0.79  \\
 BSDL~324 & 0.80    & 1.11  &0.73  \\
 BSDL~270 & 2.45    & 3.64  &0.67  \\
 BCDSP~1  & 0.78    & 1.19  &0.66  \\
 
\hline
\end{tabular}
\end{center}
\label{Q_val2.tab}
\end{table}

\begin{table}[h]
\caption{Values of Q for all the  stars in quadrilateral areas corresponding to the CCD fields (see text for details).}
\begin{center}
\begin{tabular}{ l c c c}
\hline
 Object&$\bar{m}$   & $\bar{s}$ &  $\textsl{Q}$ \\
\hline
  NGC~1763 & 0.68  &0.98   &  0.70 \\
  NGC~1769 & 0.70  & 1.13  &  0.62\\
  LH~9 East  & 0.71  & 0.97  &  0.74\\
 (NGC~1761 E) &&&\\
  LH~9 North & 0.71  &  0.97  &  0.74\\
(NGC~1761 W,HD~32228,&&&\\
BCDSP~1,BSDL~270) &&&\\
  LH~9 South & 0.71  & 0.97   &  0.74\\
  (NGC~1760) &&&\\
Field stars&0.76 & 1.12 &0.67\\
\hline
\end{tabular}
\end{center}
\label{Q_val_ccd.tab}
\end{table}

\begin{table}[h]
\caption{Values of Q for PMS (fainter than F435W $\sim 20$ ) in quadrilateral areas corresponding to the CCD regions.}
\begin{center}
\begin{tabular}{ l c c c}
\hline
Object  & $\bar{m}$   & $\bar{s}$ &  $\textsl{Q}$ \\
\hline
  NGC~1763  &  0.70 &  1.00 &  0.70 \\
  NGC~1769  &  0.71 &  1.15 &  0.62 \\
  LH~9 East  &  0.77 &  1.11 &  0.70 \\
  (NGC~1761 E)&&&\\
  LH~9 North &  0.75 &  1.04 &  0.72 \\
  (NGC~1761 W,HD~32228, &&&\\
BCDSP~1,BSDL~270) &&&\\
   LH~9 South &  0.75 &  1.12 &  0.67 \\
  (NGC~1760) &&&\\
\hline
\end{tabular}
\end{center}
\label{Q_val_ccd_r.tab}
\end{table}

\begin{table}[!bh]
\caption{Values of Q for BMS (F435W$<$20) stars (quadrilateral areas).}
\begin{center}
\begin{tabular}{ l c c c}
\hline
Object  & $\bar{m}$   & $\bar{s}$ &  $\textsl{Q}$ \\
\hline
  NGC~1763  &  0.63 &  0.93 &  0.68 \\
  NGC~1769  &  0.67 &  1.12 &  0.58 \\
  LH~9 East  &  0.76 &  1.09 &  0.69 \\
  (NGC~1761 E)&&&\\
  LH~9 North &  0.70 &  0.97 &  0.72 \\
  (NGC~1761 W,HD~32228, &&&\\
BCDSP~1,BSDL~270) &&&\\
   LH~9 South &  0.76 &  1.16 &  0.65 \\
  (NGC~1760) &&&\\
\hline
\end{tabular}
\end{center}
\label{Q_val_ccd_b.tab}
\end{table}

\begin{figure*}[!th]
\centering
\parbox{10cm}{
\resizebox{10cm}{!}{\includegraphics{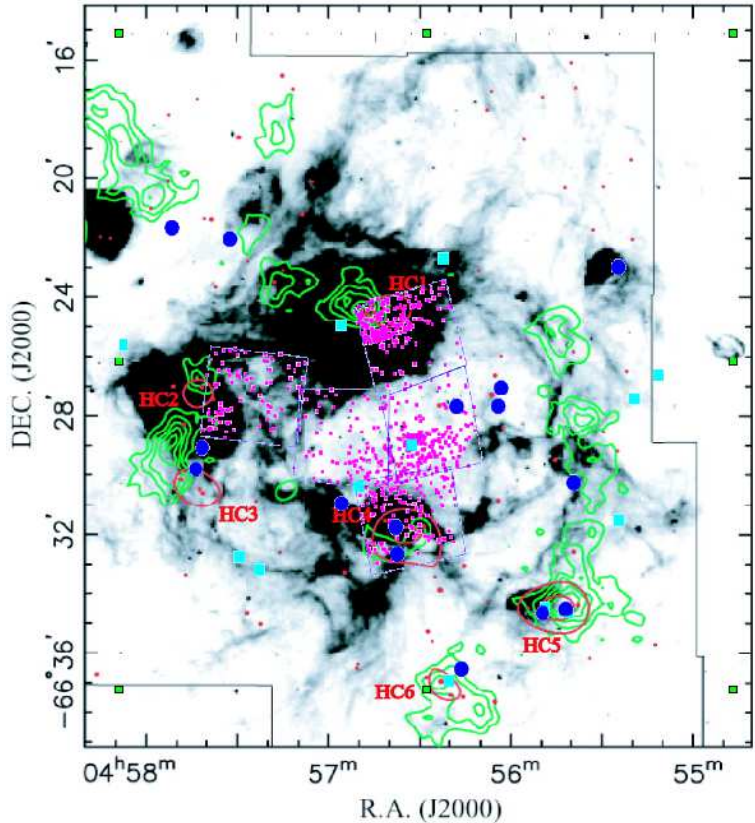}}
}
\vspace{1truecm}
\caption{H$\alpha$ emission map (inverted intensity scale) by McLow et al (1998). Large boxes show the ACS/WFC fields.  Green contours indicate the CO clouds (Israel et al 2003),  small red dots show the PMS candidates selected on the basis of ACS photometry, large green squares show the location of YSOs type II selected using IR Spitzer photometry, large blue dots represent the location of YSOs type I. Finally red circles (marked by HC1 to HC6) show the Herbig Ae/Be candidates by \cite{2006AJ....132.2653H}.}
\label{ima1}
\end{figure*}

\item OB  candidate distribution has \textsl{Q}= 0.85. This indicates  a diffuse population. Two  main concentrations are found however,  associated to HD~32228 (and in general to NGC~1761) and NGC~1763. In these objects, the positions of  the concentration of OB stars are very close to the main concentration of PMS stars.

\item Inside N11 no significant difference is found in the clustering degree of brighter blue main sequence stars (F$435W \leq 20$) and fainter red stars (F435W $>$  20) which suggests that they are formed in the same process.
A closer look reveals that the regions present a different behavior concerning the clustering degree as a function of magnitude (see Fig.~\ref{pmscum.fig}). While in  NGC~1763, LH~9 East and North bright stars are more  concentrated,  in NGC~1769 ad LH~9 South are quite dispersed. PMS fainter than F435W $\sim 20$  do not show any trend with regard to the magnitude.

\item Since young clusters are expected to evolve from hierarchical configurations to  more concentrated ones \citep{2009ApJ...694..367S}, the fact that there is a  range of structures with a  scatter in  \textsl{Q} may indicate that the objects are at different evolutionary stages.

\begin{figure}[!h]
\centering
\resizebox{7.5cm}{!}{\includegraphics{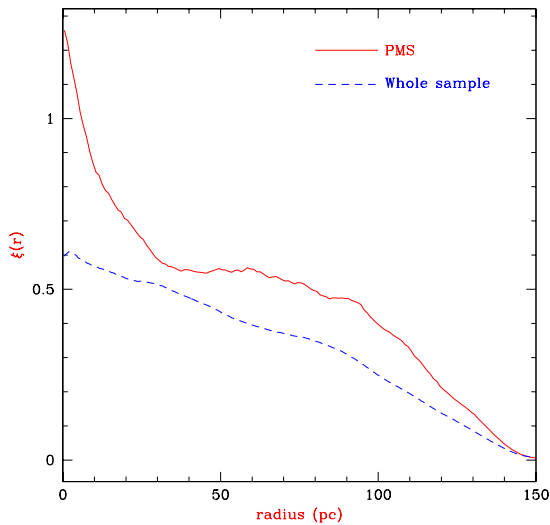}}
\caption{Two-point correlation function of the PMSs (solid line) and of the whole sample of stars (dashed line). }
\label{autocross}
\end{figure}

\begin{figure*}
\parbox{5.8cm}{
\resizebox{5.8cm}{!}{\includegraphics{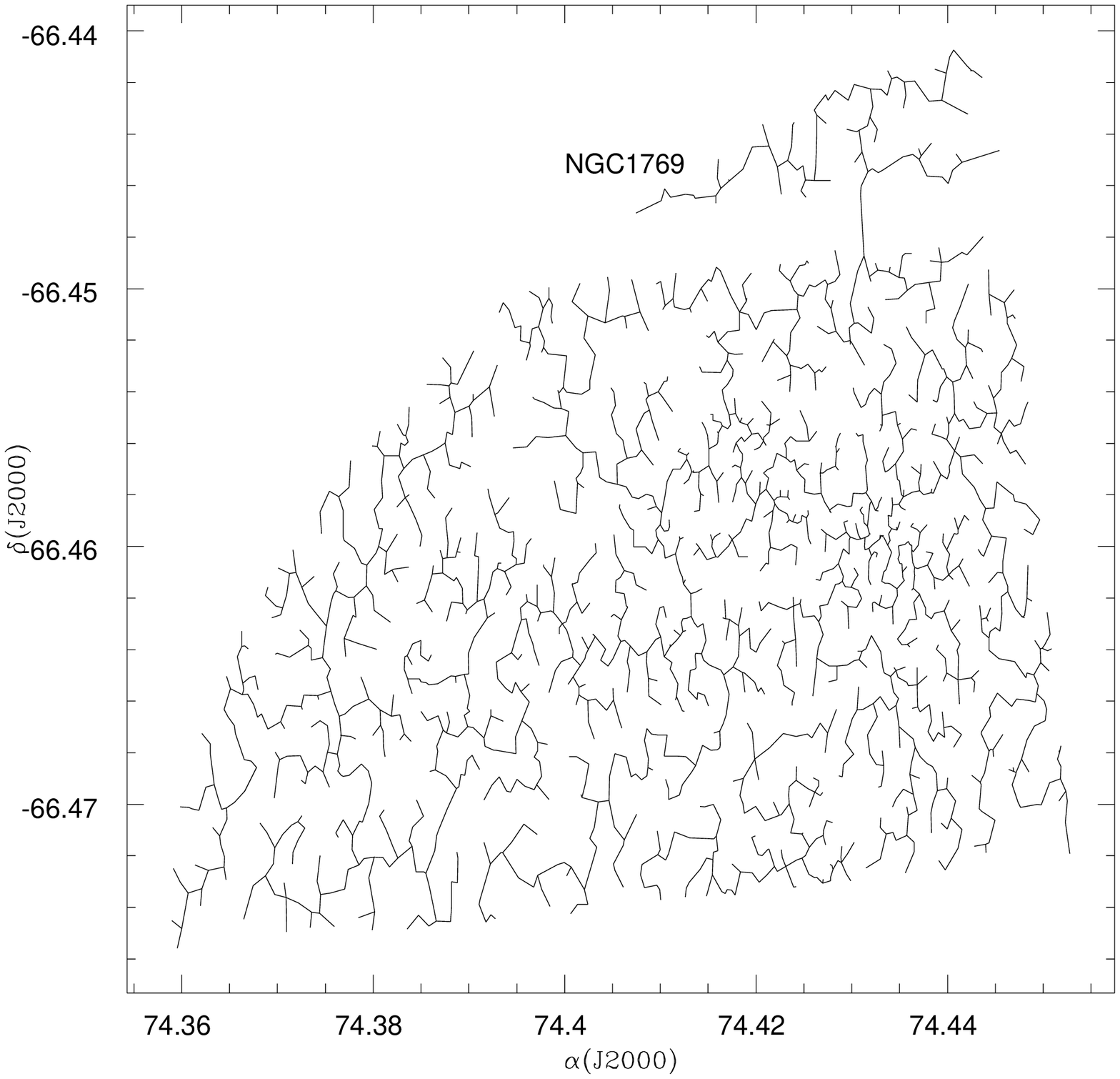}}\\
}\hspace{0.2cm}
\parbox{5.8cm}{
\resizebox{5.8cm}{!}{\includegraphics{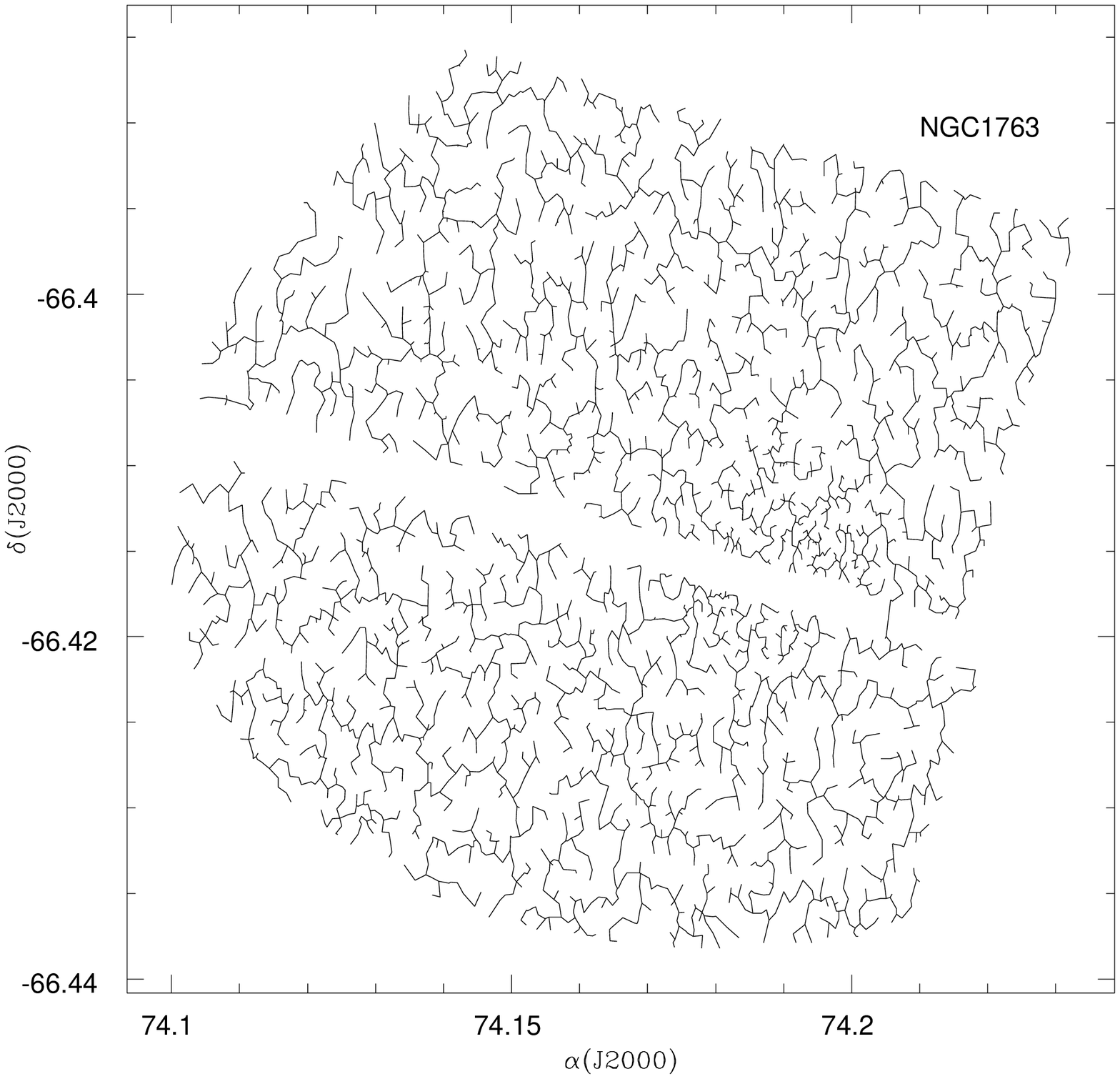}}\\
}\hspace{0.2cm}
\parbox{5.8cm}{
\resizebox{5.8cm}{!}{\includegraphics{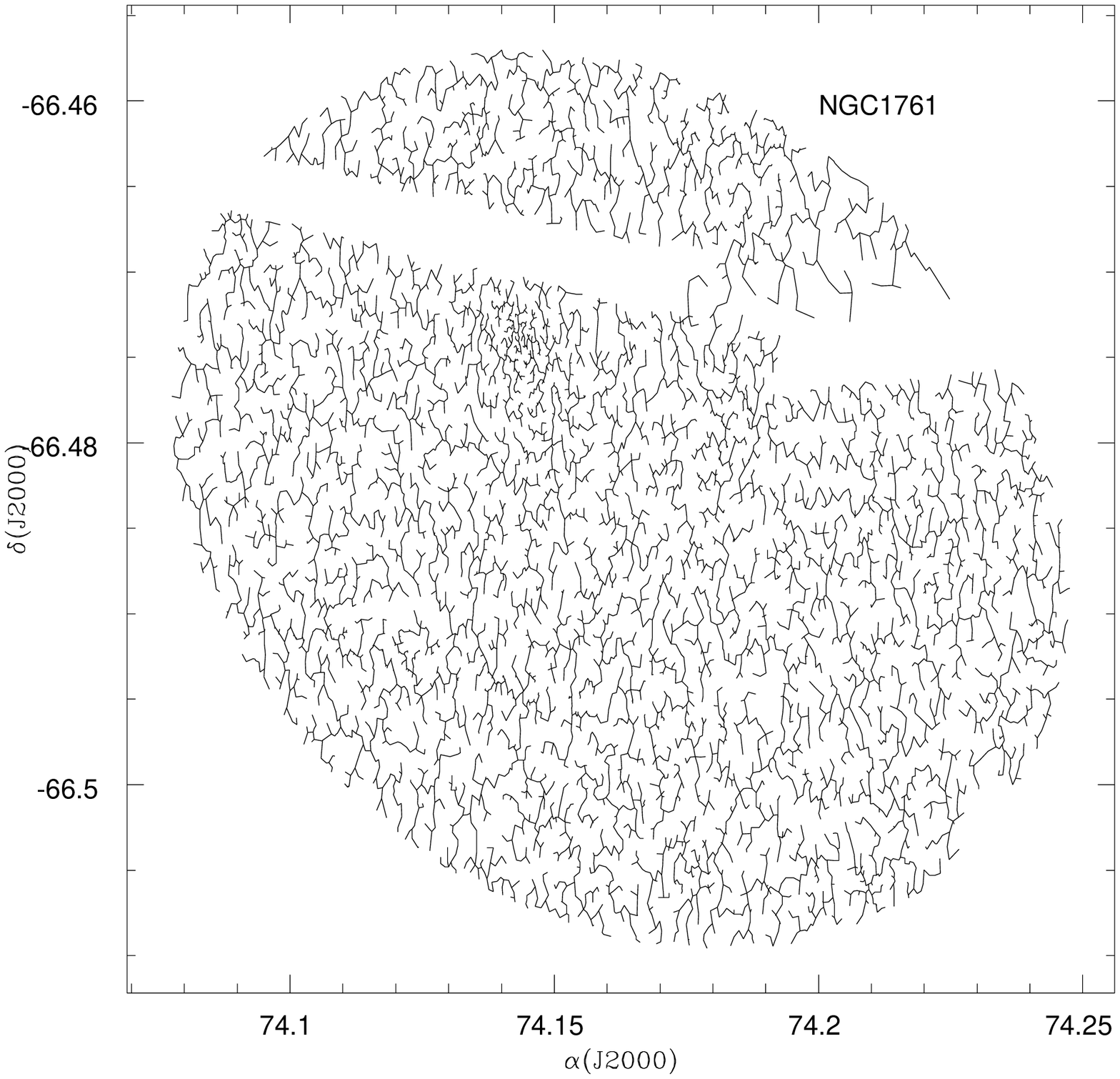}}\\
}
\\
\parbox{5.8cm}{
\resizebox{5.8cm}{!}{\includegraphics{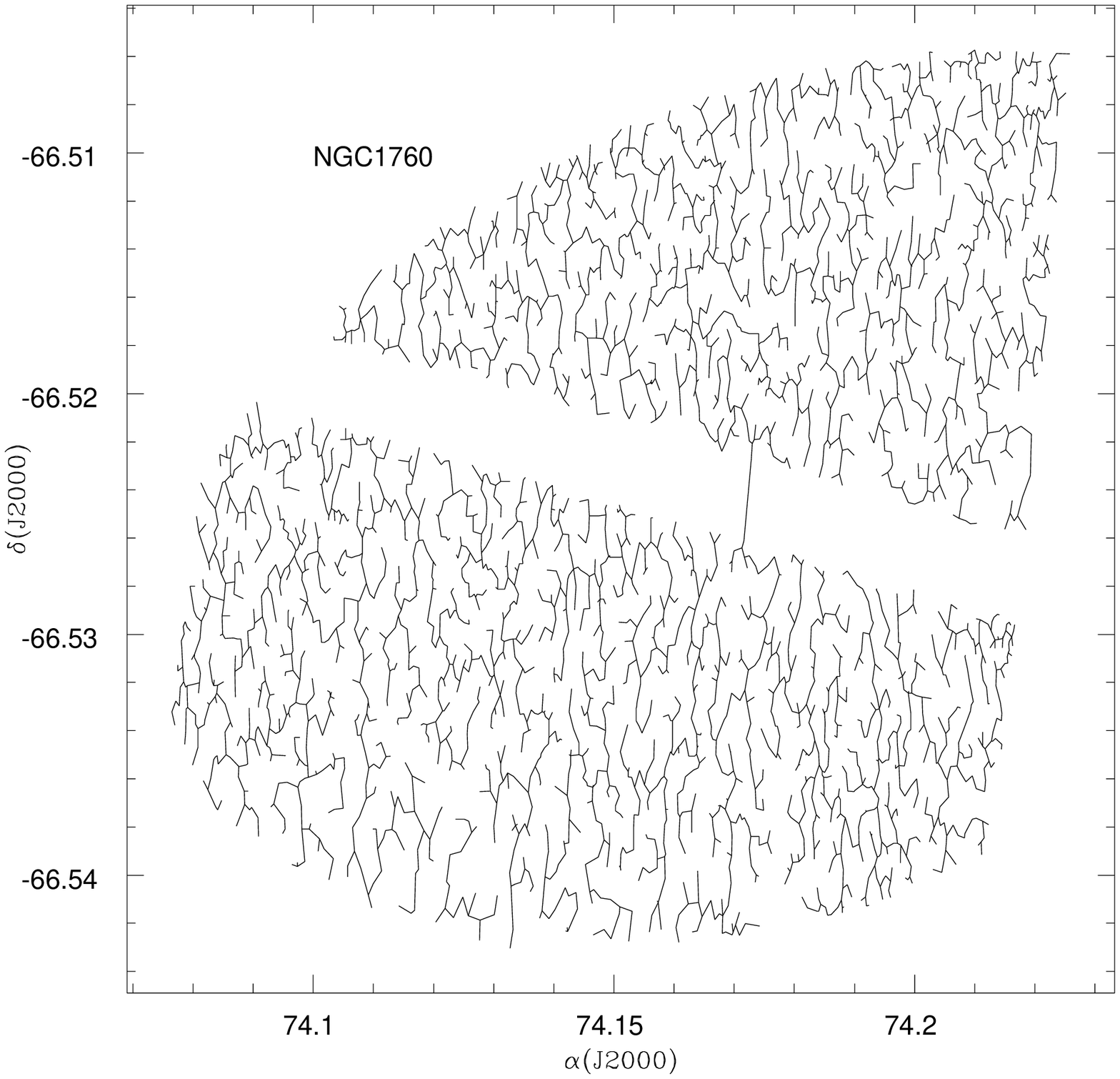}}\\
}\hspace{0.2cm}
\parbox{5.8cm}{
\resizebox{5.8cm}{!}{\includegraphics{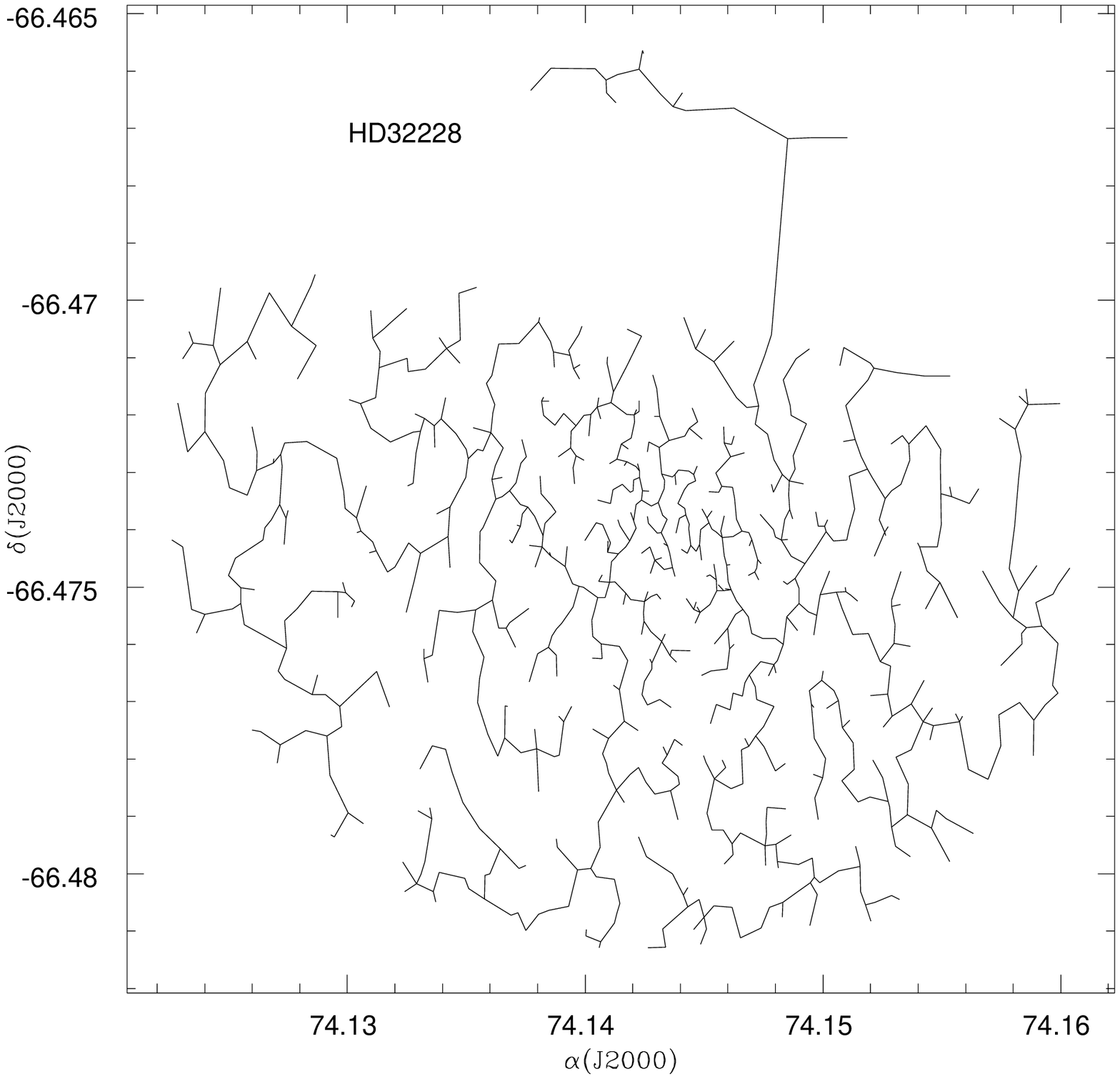}}\\
}\hspace{0.2cm}
\parbox{5.8cm}{
\resizebox{5.8cm}{!}{\includegraphics{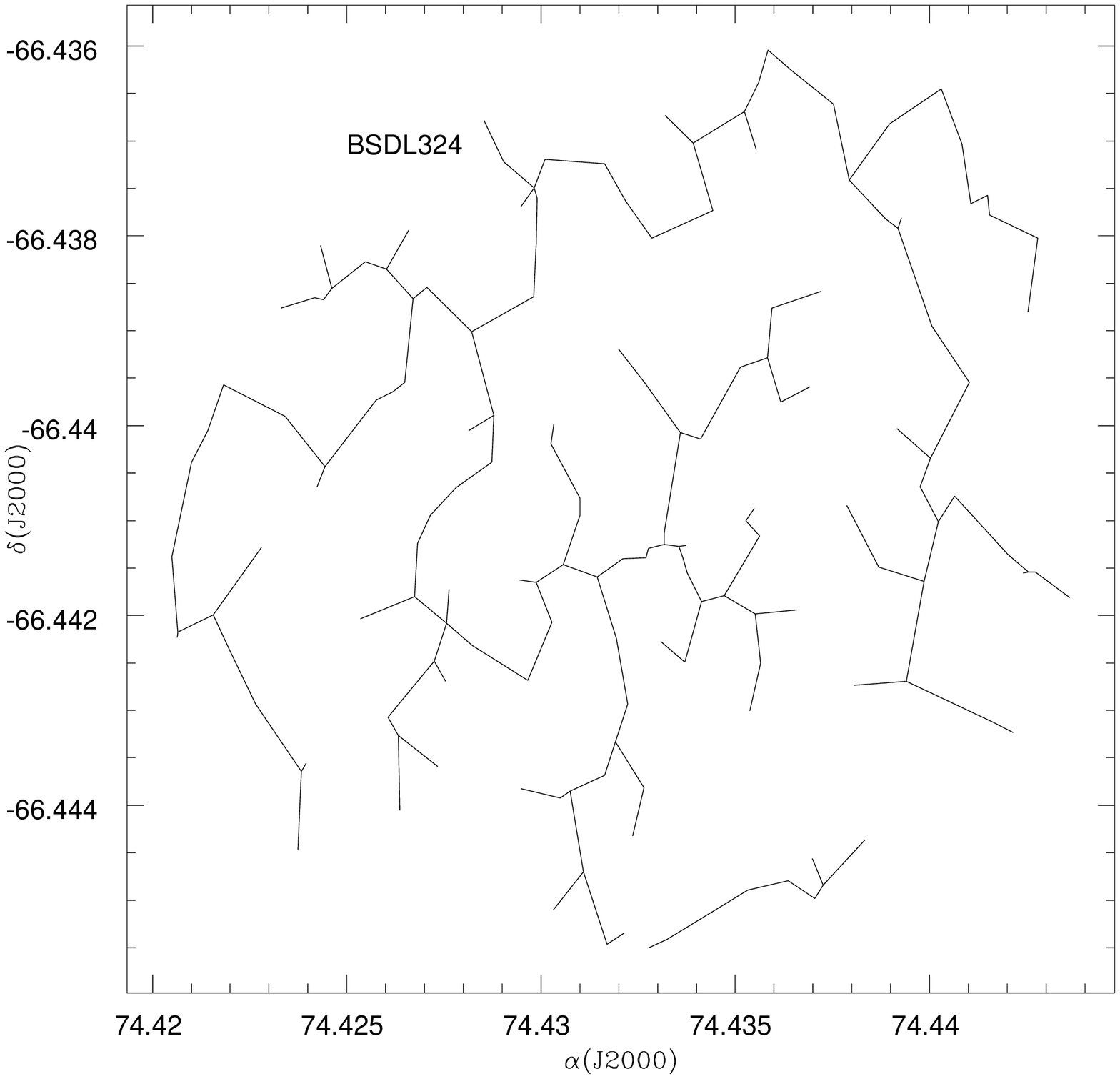}}\\
}
\\
\parbox{5.8cm}{
\resizebox{5.8cm}{!}{\includegraphics{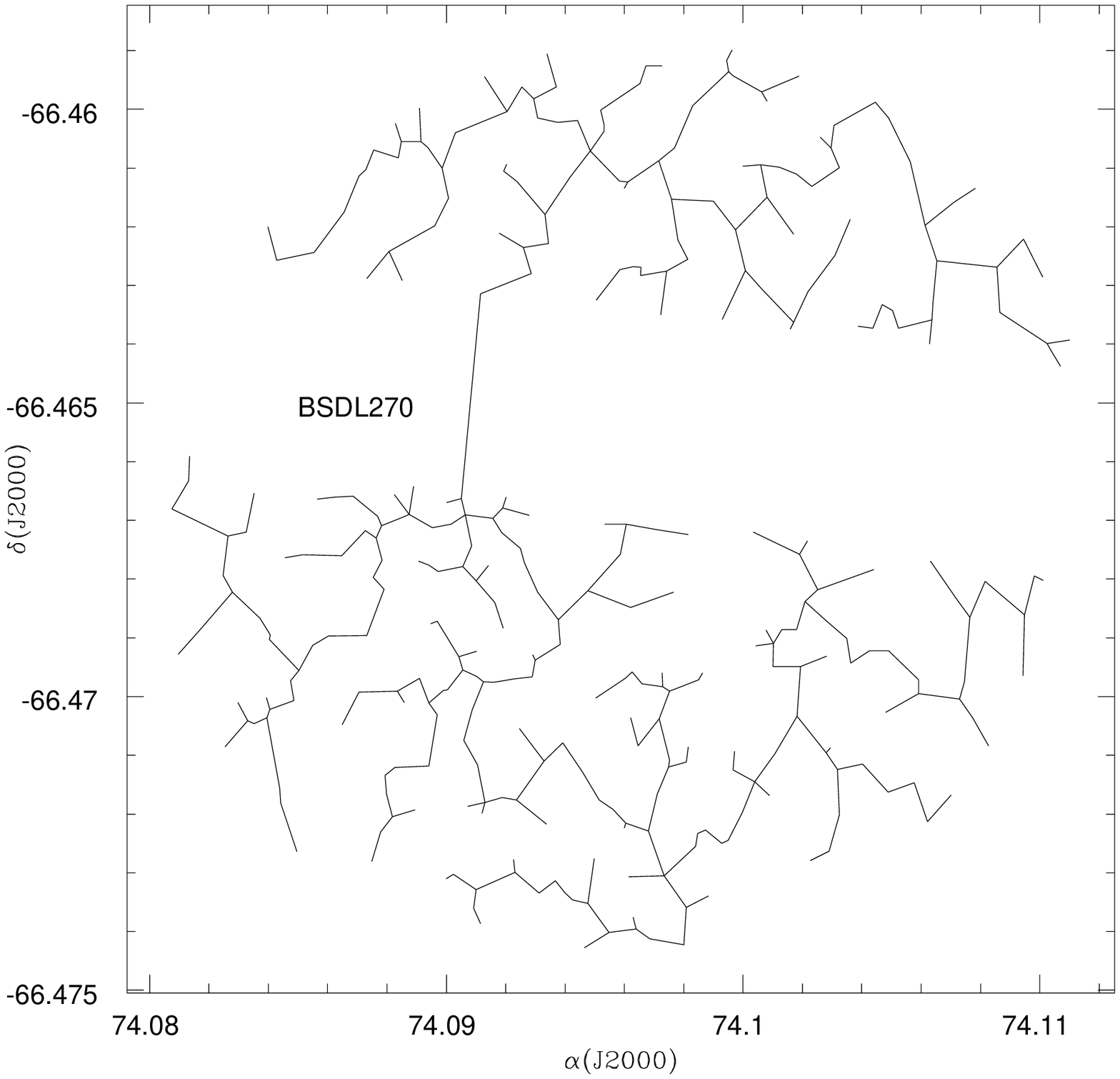}}\\
}\hspace{0.2cm}
\parbox{5.8cm}{
\resizebox{5.8cm}{!}{\includegraphics{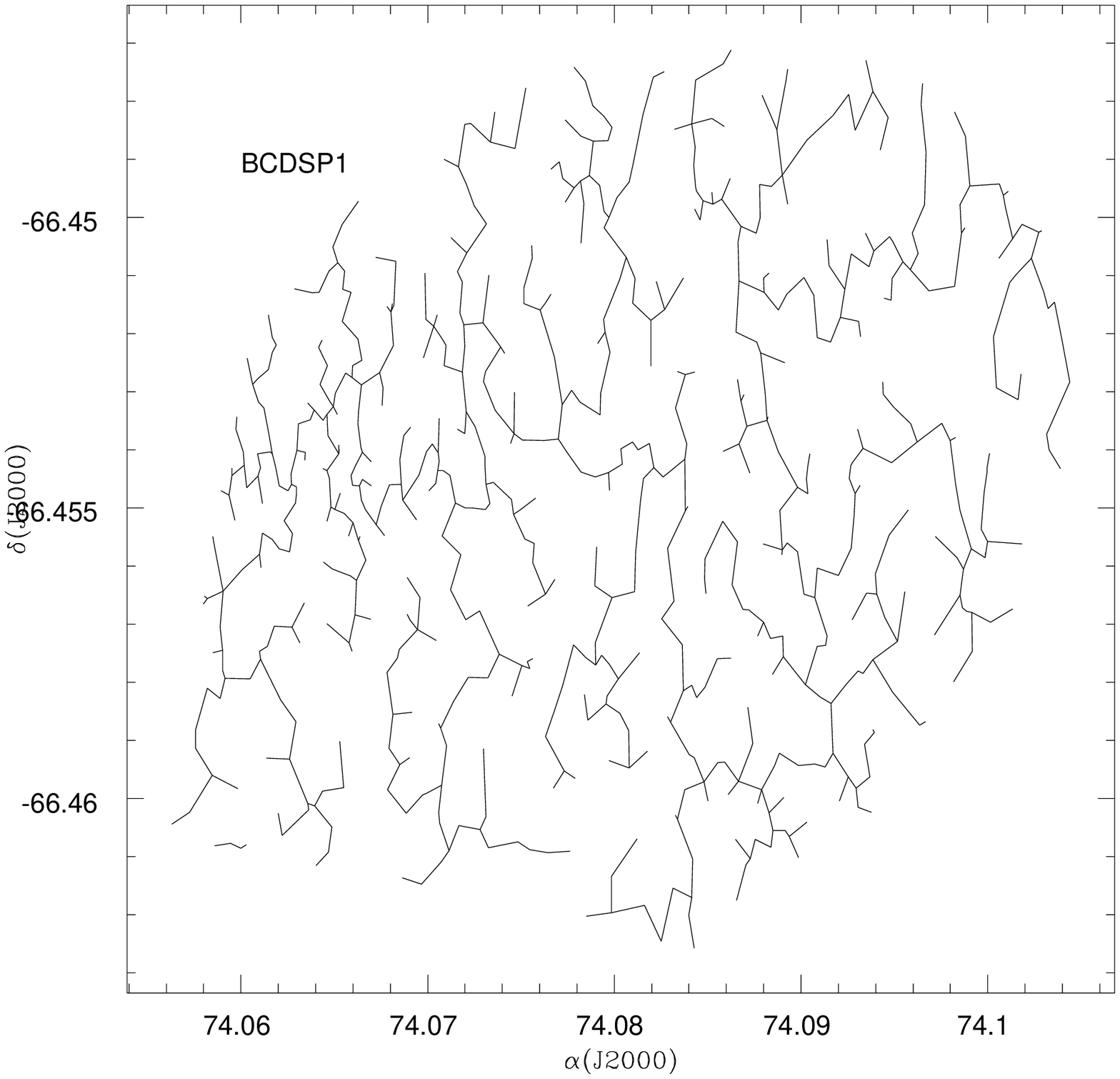}}\\
}\hspace{0.2cm}
\parbox{5.8cm}{
\resizebox{5.8cm}{!}{\includegraphics{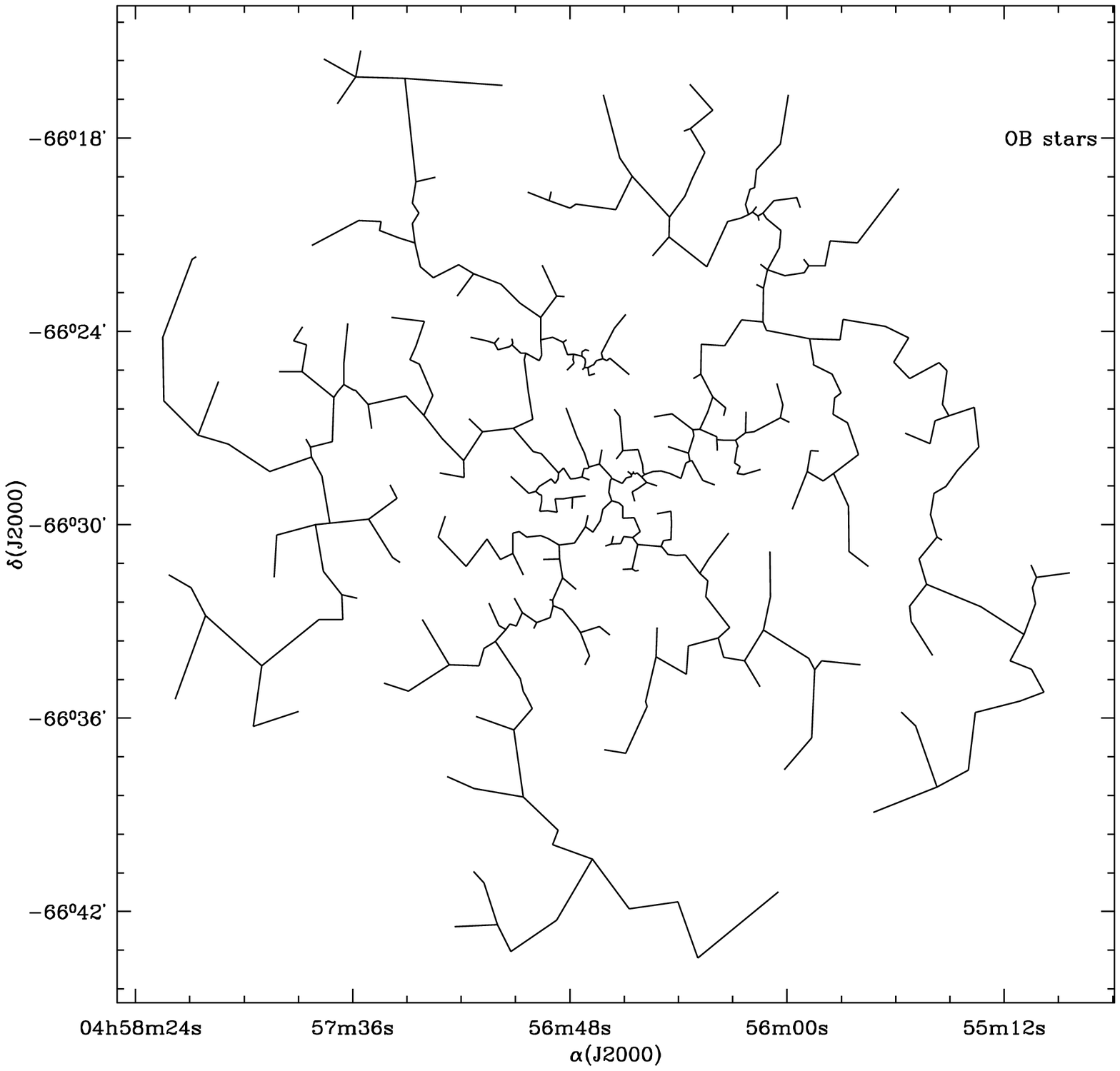}}\\
}

 \caption{The MST of the regions of NGC~1769,NGC~1763,NGC~1761, NGC~1760,
 HD~32228, BSDL~324, BSDL~270 and BCDSP~1 and for OB star candidates.}
\label{NGC_grafo.fig}
\end{figure*}

\item  \textsl{Q} is found to have no dependence on the  age of the clusters, even if the  oldest objects HD~32228 and in general LH~9 North have a slightly higher \textsl{Q}. It should be remembered that this comparison is complicated by the presence of  a large age dispersion detected in the fields.

\item  This is at odds with what we find in the star forming region in our Galaxy, where  \textsl{Q} correlates with the age.  \citet{2009ApJ...694..367S} find similar behavior  in the region of NGC~346 in the SMC and suggest that this can be explained as the footprints of inhomogeneity in the original turbulent cloud. In fact \textsl{Q} is expected to depend on a variety of additional  factors like the initial conditions or turbulent energy in the clouds.  

\item The value of \textsl{Q} itself in the field population is rather low, around 0.67, meaning that the field still has substructures due to the original turbulent environment or showing the result of random motion configurations. 
The  value we find is compatible with  \textsl{Q}=0.7 derived by \citet{2009MNRAS.392..868B} as a reference value for the field of the LMC.

\end{itemize}

\newpage
\section{Conclusions}
\label{conclu}

Young stars are well known to  trigger star formation on a local scale by injecting energy in the surrounding interstellar medium.   N~11 region in the LMC is often presented as one of the most promising candidates of  a triggered star formation.
 In this paper we discussed the star formation process in N~11. We reported on the discovery of PMS and YSO candidates associated with N~11 from HST ACS/WFC photometry  and archive Spitzer data. While HST observations can give information about faint, exposed pre-main sequence candidates, IR data allowed us to detect embedded young stellar objects. The main conclusions are as follows:

\begin{itemize}

\item
A large population of PMSs  is found in this region.
 The comparison with the isochrones  shows that PMSs have masses from 1.3 ~M$_\odot$ to 2.0  ~M$_\odot$, depending on their ages. YSOs (type I and II) having ages $<1 $ Myr are found at the same location as the  PMSs.

\item
 The analysis of HST data implies that PMSs in LH~9 are  slightly older than in the other regions of N~11. A prolonged star formation is detected in LH~9 South from 2 to 10 Myr. However, the majority of the stars are older than 6 Myr, while LH~9 North, LH~10(center) and LH~13 seem to be younger (having ages of 5, 2, 2 Myr respectively). A large age spread from 1 to 6 Myr is found from the spectroscopy of blue stars in these regions. Remarkably, the present age determinations of the PMSs agree (inside the errors) with the ages derived from literature spectroscopy of blue bright stars.

\item The presence of YSOs  in LH~9 South, LH~13 (type I) and LH~9 (North), LH~10 (Type II)  suggests that the star formation in the region is still going on. The most conspicuous concentrations of PMSs correspond to the location of OB stars. Concerning Herbig Ae/Be stars, they are found in all these locations, confirming the presence of a population with ages in the range of 1$-$3 Myr. All the indicators present a coherent picture of the star formation in the region.

\item PMSs  present a higher degree of clustering in comparison with
the whole star sample.
The distribution  of the nearest neighbor distance  reflects the structure of the interstellar medium. The full width half maximum of the peak of the correlation function for PMSs is  $\sim 10 $ pc, which is  the order of the typical size of molecular clouds in the SMC and LMC, which extends from about 10 to 100 pc.

\item  In all the fields related to N11, the values of \textsl{Q} $\leq$  0.8 are typical of the products of hierarchical star formation. Clusters are expected to evolve from hierarchical structures to more concentrated ones. Different clusters/associations show slightly different values of  \textsl{Q}, suggesting that they might be in slightly different evolutionary stages. Inside N11, no significant difference is found in the clustering degree of brighter blue main sequence stars (F$435W \leq 20$) and fainter red stars (F435W $>$  20), suggesting that they might be part of the same formation process.

\item \textsl{Q} is found to have no dependence on the  age of the clusters, even if the  oldest objects HD~32228 and in general LH~9 North have a slightly higher \textsl{Q}.
 These  results disagree with the behavior found in our Galaxy. These results are quite difficult to interpret. They can be explained as the footprints of  inhomogeneity in the original turbulent cloud. It should be kept in mind that this comparison is complicated by the presence of  a large age dispersion detected in the fields.
\end{itemize}

\bibliographystyle{apj}
\bibliography{lmc}

\end{document}